\documentclass[twocolumn,preprintnumbers,amsmath,amssymb,superscriptaddress,graphicx,graphics,pstricks,prd]{revtex4-2}
\usepackage{hyperref}
\usepackage{amsmath}
\usepackage[x11names]{xcolor}
\usepackage{graphicx}
\usepackage[utf8]{inputenc}
\usepackage{pstricks}
\usepackage{pst-node}
\usepackage{pst-coil}
\usepackage{pst-grad}
\usepackage{multido}
\usepackage{dcolumn}
\usepackage{bm}
\usepackage{physics}
\usepackage[normalem]{ulem}

\usepackage{soul}

\begin{document}

\title{Dispersive interaction between two atoms in Proca Quantum Electrodynamics}

\author{Gabriel Camacho de Pinho, Carlos Augusto Domingues Zarro, Carlos Farina}
\affiliation{Instituto de Fisica, Universidade Federal do Rio de Janeiro, Av. Athos da Silveira Ramos, 149,
Centro de Tecnologia - bloco A, Cidade Universitária, Rio de Janeiro - RJ - CEP: 21941-909, Brasil}

\author{Reinaldo de Melo e Souza}
\affiliation{Instituto de Física, Universidade Federal Fluminense, Av. Litorânea, s/n, Niter\'{o}i,
RJ, 24210-346, Brasil}

\author{Maur\'{\i}cio Hippert}
\affiliation{Illinois Center for Advanced Studies of the Universe \& Department of Physics, University of Illinois Urbana-Champaign, Urbana, IL 61801, USA}
\affiliation{Instituto de Física, Universidade do Estado do Rio de Janeiro, Rua São Francisco Xavier, 524, Rio de Janeiro, RJ, 20550-013, Brasil}

\begin{abstract}
We analyze the influence of a massive photon in the dispersive interaction between two atoms in their fundamental states.  We work in the context of Proca Quantum Electrodynamics. The photon mass not only introduces a new length scale but also gives rise to a longitudinal polarization for the electromagnetic field. We obtain explicitly the interaction energy between the atoms for any distance regime and consider several particular cases. We show that, for a given interatomic distance, the greater the photon mass the better it is the non-retarded approximation.
 \end{abstract}


\maketitle
%
%
%
%
\section{Introduction}

Neutral but polarizable atoms with no permanent multipoles placed in vacuum will still interact with each other. This can be understood from the Heisenberg uncertainty principle which enforces dipole fluctuation even in states where the
expectation value of the dipole operator vanishes. These
forces are the so-called dispersive interactions
\cite{milonni1994quantum,israelachvili,buhmann1,buhmann2,milton}. They are responsible for several phenomena not only in Physics, but also in Chemistry, Biology and many other areas of science. For instance, they explain why noble gases condensate. Besides, since dispersive forces strongly depend on the polarizabilities of the atoms/molecules involved, which in turn scale with the volumes of  such atoms/molecules, these forces explain why the temperatures of condensation of noble gases are greater for the larger atoms (with Helium having the smallest and so on). By the same argument, dispersive forces may become dominant in macromolecules interactions.  Increase attention are being given to these forces since they  are responsible for the cohesion between different layers of the van der Waals heterostructures, materials obtained by  stacking different 2d materials and which display fascinating properties\cite{Geim2013}. There is still an intense research going also from the quantum field theory perspective\cite{castillo2024casimir,castillo2022casimir,fosco2024casimir,fosco2020casimir,fosco2016casimir,romaniega2023casimir,cavero2021casimir,shajesh2016casimir,shajesh2017casimir,li2022casimir,milton2017casimir,milton2019remarks}. In Biology, they are responsible for the softness of our skin and explain the remarkable adhesion of geckos on the walls of our houses\cite{gecko}. We can go even further with more bizarre examples, like the important role of dispersive forces in the generation of the high electrostatic potentials ocurring during the storms, among others \cite{lamoreaux2007casimir,scheeres2010scaling}. Differently from other kinds of intermolecular forces, dispersion forces are always present due to the ubiquity of Heisenberg uncertainty principle.

 We can think of dispersive forces as a consequence of the exchange of virtual photons between the atoms.
 In Maxwell electrodynamics in vacuum, photons are massless, and this is responsible for some important features of these forces, such as the long range character of the interaction.
 However, under certain circumstances, photons may acquire an effective mass due to shielding mechanism. One example arises in colloidal systems. When two atoms interact within an electrolyte solution, the potential generated by the atoms ionize the solution and when the potential is not very strong - that is if the atoms are not very close to each other - we may treat the interaction within the linearized Poisson-Boltzmann equation\cite{mahanty1976}. In this scenario the Maxwell field in the solution behaves as a free Proca effective electromagnetic field, with the photon mass being given by the inverse of the Debye wavelength.
 Similar mechanisms arise in metals, or inside waveguides. Indeed, in many situations, a gas of photons between conducting plates may be thought of as a massive bi-dimensional photon gas \cite{chiao1999bogoliubov}.
 Such an effective photon mass affects not only real, but also virtual photons, and thus will have consequences for dispersive interactions.
 In fact, a recent work \cite{Souza2015} has also shown that the dispersive interaction between two
atoms inside a plane capacitor is screened due to non-additive effects, in a way that can be effectively described by endowing photons with a
mass inversely proportional to the separation between capacitor plates. Also, electric as well as magnetic condensates may lead to effectively massive gauge fields and can be used to investigate superconductivity, confinement in QCD, and an holographic duality scenario to explain the metal-insulator transition in condensed matter \cite{Grigorio:2012jt,Guimaraes:2012tx,Reinosa:2024njc,Rougemont:2015gia}.

Even in vacuum, a tiny, yet non-vanishing, photon mass is also present in scenarios of physics beyond the standard model \cite{Pospelov:2008zw,Spallicci:2020diu}. Moreover, these Yukawa-type corrections appear in axion physics and modifications of Newtonian gravity. Casimir effect experiments can be used to impose bounds on these Yukawa parameters\cite{bordag1998constraints,fischbach2001new,mostepanenko2016progress}.
However small it may be, a finite mass for the photon
changes a lot our physical picture of the world. Indeed,
as pointed out in \cite{adelberger2007photon}, the above bounds may
not be correct depending upon the microscopical
origin of the mass. For example, if it appears from
the Higgs mechanism, it is possible for large-scale fields
to be effectively Maxwellian, and in this case astrophysical observations will be insensitive to
the mass of the photon.

Many different observations place very stringent bounds for the photon mass.
The most rigorous restrictions come from astrophysical observations, which put an upper bound of $10^{-27}$\, eV to the mass of the photon \cite{chibisov1976astrophysical}.
Particle properties, such as the anomalous magnetic moment of the electron, also place strict limits on its mass \cite{accioly2010upper}.
Interestingly, some mass bound for photons can be obtained using Schuman ressonances \cite{malta2022constraining}.
In Ref.~\cite{malta2023shining} another terrestrial experiment was proposed, in which the longitudinal mode of photons would be measured to restrict the mass of the photon.
Two comprehensive reviews, written by the same authors almost 40 years apart, can be found in \cite{goldhaber1971terrestrial,goldhaber2010photon}.

Ultimately, dispersive interactions akin to van der Waals forces are to be expected in any theory with Abelian vector bosons, as is the case in some models of nuclear physics. In that context, vector mesons have a considerable mass in vacuum, which are not protected by the global symmetries linked to these particles.
In particular, short-range repulsion between nucleons due to the exchange of massive omega vector mesons plays an important role in the structure of both nuclei and neutron stars \cite{norman1997compact,Horowitz:2000xj,Chen:2014sca}.
In fact, the interplay between attractive interactions mediated by scalar mesons and repulsive interactions, mediated by such vector mesons, are thought to be the mechanism behind nuclear saturation \cite{Walecka:1974qa,Boguta:1982wr}.
The omega meson couples to baryon charge, while mesons in general --- composed of a valence quark-antiquark pair --- should display fluctuations of the dipole moment of baryon number density.
Therefore, omega meson exchange should lead to dispersive forces between mesons. Recently, the coupling of omega mesons to pions and sigmas has been proposed as a mechanism to explain the speed of sound peak expected in ultra-dense nuclear matter \cite{Pisarski:2021aoz}.
Dispersive van der Waals forces within the context of Quantum Chromodynamics have been previously explored in the literature  \cite{willey1978quark,Fujii:1999xn}. 

 The simplest description of massive vector mesons is given by Proca electrodynamics, in which gauge invariance is explicitly lost \cite{proca1936theorie}.
While, in some applications, a more fundamental description  may be provided by other formulations, which might preserve gauge symmetry, such as the Higgs mechanism or the Stueckelberg action, Proca electrodynamics can be motivated as a simplified version of these formulations, obtained, for instance, by treating extra fields (e.g., the Higgs field) as static, uniform backgrounds. The investigation of dispersive forces mediated by massive photons also raises interesting conceptual questions. For instance, what is the nature of the limit in which the photon mass is taken to zero?  The disappearance of the longitudinal polarization could raise the possibility of a discontinuous transition to Maxwell theory, whereas a continuous transition seems to always be found. Yet, the precise dependence of van der Waal forces on the mediator mass near this limit will be revealing of how sensitive these forces are to small photon mass in scenarios beyond the standard model. In the context of the Casimir effect this discussion has revealed surprising subtleties. It was initially predicted that the zero mass limit would be singular due to the longitudinal modes\cite{daviesPLB}, which was subsequently disproved\cite{barton1984casimir,barton1985casimir} by showing that in the zero mass limit these modes decouple from matter and the plate becomes transparent to them. In this way only transversal modes contribute to the Casimir attraction between plates assuring that this limit is well behaved.
Furthermore, understanding dispersive interactions mediated by massive photons may also furnish new insights into Maxwell electrodynamics. For example, a photon mass breaks the degeneracy between the group, phase and signal propagation velocities, present in standard electrodynamics. This will enable us to verify which one plays the dominant role in the physics of dispersive forces.


 In this work we employ Proca electrodynamics to investigate the effects of a finite photon mass on the dispersive interaction between two atoms. This subject has recently got renewed interest in the search for signatures of physics beyond standard models\cite{mattioli2019casimir}.  In this paper we employ a Hamiltonian originally proposed by P. Milonni\cite{milonni1994quantum} and show that it is possible to analyze the Proca dispersive interaction within first order perturbation theory. Afterwards we analyze the short distance limit, known as the non-retarded regime. We study this limit by two complementary approaches. In section II we evaluate the interaction energy without quantizing the electromagnetic field, while in section III we re-obtain the results of section II by taking the appropriate limit of the full quantum electrodynamics treatment. By doing so, we are able to address the question of when it is necessary to quantize the electromagnetic showing that it leads to a weaker condition for the interatomic distance. In other words, a photonic mass extends the range of validity of the non-retarded regime. We leave section IV for our final remarks.

\section{London-Proca interaction\label{Non-retarded regime}}

Let us consider two atoms, $A$ and $B$, held at positions $\mathbf{r_A}$ and $\mathbf{r_B}$, respectively and for convenience let us define $\boldsymbol{R}=\boldsymbol{r}_B-\boldsymbol{r}_A$.  Throughout this paper we shall assume the magnitude of the distance $R$ to be much larger than the typical size of the atoms,  enabling us to employ the electric dipole approximation. Hence, dispersive forces arise from correlations between the fluctuating dipoles in each atom. In the London regime, we consider the field generated by the atomic dipoles as electrostatic. This is a good approximation when the atoms are close to each other, and in the next section we employ a full quantum electrodynamics calculation in order to establish the precise condition $R$ must satify.
In the dipole approximation,
the interaction Hamiltonian describing the coupling
between the atoms is given
by
\begin{equation}
H_{int}=-\mathbf{d}_A\cdot\mathbf{E}_B(\mathbf{r_A}) \, , \label{hint}
\end{equation}
where $\mathbf{d}_A$ denotes the electric dipole
operator of atom $A$ and $\mathbf{E}_B(\mathbf{r_A})$
stands for the electrostatic field
created by the dipole $B$ at the position
of atom $A$. The electrostatic field $\mathbf{E}_B$ depends upon
the dipole operator of atom $B$, $\mathbf{d}_B$, and is given by (see appendix \ref{dipoleProca})
\begin{eqnarray}
\mathbf{E}_B(\mathbf{r}_A)&=&-\frac{e^{-\mu R}}{R^3}(\mu R+1)\mathbf{d}_B+\cr\cr
&+&\frac{e^{-\mu R}}{R^5}(\mu^2R^2+3\mu R+3)(\mathbf{d}_B\cdot\mathbf{R}))\mathbf{R} \, , \label{estatico}
\end{eqnarray}
where $\mu \equiv mc\,(\hbar)^{-1} > 0$,  with $m$ denoting the photon mass in Proca electrodynamics. Substituting the last equation into the Hamiltonian (\ref{hint})
we obtain
\begin{eqnarray}
&&H_{int}=\frac{e^{-\mu R}}{R^3}(\mu R+1)\mathbf{d}_A\cdot\mathbf{d}_B+\cr\cr
&-&\frac{e^{-\mu R}}{R^5}(\mu^2R^2+3\mu R+3)(\mathbf{d}_A\cdot\mathbf{R})(\mathbf{d}_B\cdot\mathbf{R}) \, . \label{hint2}
\end{eqnarray}
Notice that this Hamiltonian is symmetric under the exchange $A\leftrightarrow B$ as it should - the interaction does not depend on which atom we choose as the source of the electric field in Eq.~(\ref{hint}).
 Denoting the ground state of atom $j$ by $|0_{j}\rangle$, the fact that the atom does not possess permanent electric dipole means that $\langle 0_j|\mathbf{d}_{j}|0_{j}\rangle=0$. Therefore, perturbation theory applied to the above Hamiltonian vanishes at first order. At second order we have
\begin{equation}
U_{\rm NR}=-\sum_{r\neq 0,s\neq 0}\frac{\langle 0_A0_B| H_{int}|rs\rangle\langle rs| H_{int}|0_A0_B\rangle}{\hbar (\omega_{r0}+\omega_{s0})} \label{pert2} \,
\end{equation}
where we denote the excited states of atoms $A$ and $B$ by $|r\rangle$ and $|s\rangle$, respectively. We assume that the atomic ground state is non-degenerate, as usual.
Choosing the $Z$-axis parallel to $\mathbf{R}$  we may rewrite Eq.~(\ref{hint2}) to obtain
\begin{eqnarray}
&&H_{int}=\frac{e^{-\mu R}}{R^3}\Bigg[(\mu R+1)(d^x_Ad^x_B+d^y_Ad^y_B) \cr\cr
&-&(\mu^2R^2+2\mu R+2)d^z_Ad^z_B\Bigg] \, . \label{h}
\end{eqnarray}
Substituting Eq.~(\ref{h}) into Eq.~(\ref{pert2}) we obtain
\begin{equation}
    U_{\rm NR} = -\Lambda\frac{e^{-2\mu R}}{9R^6}p(\mu R) \, , \label{nonretarded}
\end{equation}
where
\begin{equation}
    p(\xi)=6+12\xi+10\xi^2+4\xi^3+\xi^4 \label{P}
\end{equation}
and $\Lambda > 0$ is a constant depending only on the atomic internal structure, given by
\begin{equation}
    \Lambda = \sum_{r\neq 0,s\neq 0} \frac{|\mathbf{d}_A^{0r}|^2|\mathbf{d}_B^{0r}|^2}{\hbar(\omega_{r0}+\omega_{s0})} \, . \label{Lambda}
\end{equation}
In the last expression we assume an isotropic response of atoms, which enables us to make the replacement $\langle 0|d_m|I\rangle\langle I| d_n|0\rangle =|\mathbf{d}^{0I}|^2\delta_{mn}/3$ for each atom. We emphasize that, in principle, $\Lambda$ depends on $\mu$, since the photon mass also affects the fields inside the atoms. Nonetheless, assuming $\mu a\ll 1$, where $a$ is the atom size, we can  consider the atom structure to be independent of $\mu$, and suppose that $\mu$ affects only the intermolecular interaction. We shall assume this to be the case henceforth.

\begin{figure}[h!]
    \centering
    \includegraphics[width=8.6cm]{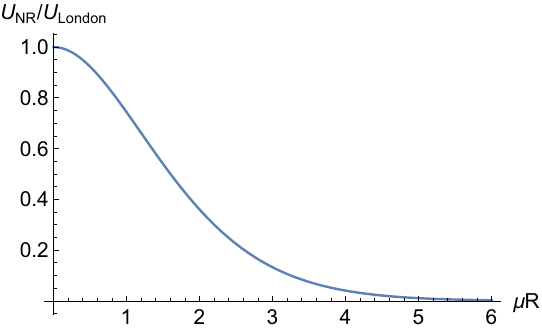}
    \caption{Interaction energy in the non-retarded regime divided by the London-van der Waals potential as a function of the dimensionless radius parameter $\mu R$.}
    \label{plot-nr}
\end{figure}

As expected, if $\mu=0$ we reobtain London's formula, hereafter denoted by $U_{\rm London}$. For $\mu R\ll 1$, the leading corrections to the London potential are given by
\begin{equation}\label{eq:asymp-nr}
    U = U_{\rm London}\left(1-\frac{1}{3}(\mu R)^2+\mathcal{O}(\mu^4R^4)\right) \, ,
\end{equation}
and hence we see that a small $\mu$ weakens the interaction, as could be expected, since the field produced by each fluctuating dipole becomes exponentially damped with distance.  Notice that the linear term is not present, which could be anticipated by Eq.(\ref{estatico}), since the Proca correction for the dipole electric field is already of order $\mu^2$, as the reader may verify.
In the regime $\mu R\gg 1$, on the other hand, the interaction between atoms is exponentially suppressed. In Fig.\ref{plot-nr} we plot the interaction energy normalized by the usual London interaction as a function of the dimensionless parameter $\mu R$. 
%

Equation~(\ref{nonretarded}) reproduces exactly the interaction energy of two atoms interacting inside a colloidal system\cite{mahanty1976}, justifying the use of Proca electrodyamics as an effective theory for these systems. Within this effective scenario the intraatomic field is Maxwellian and thus $\Lambda$ is not dependent on $\mu$. This means that the ratio $U_{\rm NR}/U_{\rm London}$ depends only on the colloidal properties which defines $\mu$ and not on the atoms partaking in the interaction.

Before closing this section, we follow \cite{craig1998molecular} (see section 7.6) and rewrite $\Lambda$ in the form
\begin{equation}
    \Lambda = \frac{9}{4\pi} \int_{-\infty}^{\infty} \alpha_A(i\chi)\alpha_B(i\chi) d\chi \, ,
\end{equation}
where $\alpha_j$ denotes the dynamical polarizability of atom $j$, defined by
\begin{equation}\label{kramers-heisenberg}
    \alpha_j (\omega) = \frac{2}{3} \sum_{I\neq 0} \frac{\omega_{I0}|\mathbf{d}_j^{0I}|^2}{\hbar(\omega_{I0}^2-\omega^2)} \, .
\end{equation}
Since the intermolecular interaction is due to the correlation between induced dipoles, the polarizability is a key concept to the physical understanding of the intermolecular interaction. Equation~\eqref{kramers-heisenberg} illustrates the dispersive character of the interaction, expressed by the dependence of the atomic response $\alpha_j$ on the frequency of the external perturbation. As we shall see in the next section, the polarizability is the most convenient starting point for analyzing the interatomic interaction within quantum electrodynamics.

\section{Quantum field theory approach\label{fullcalculation}}

\subsection{The complete potential}\label{completepotential}

When retardation effects in the electromagnetic interaction are relevant, we must include into the description of our physical system the mediator of the interaction between the atoms, namely, the electromagnetic field. And since dispersive forces have a quantum nature, we must quantize this field~\cite{milonni1994quantum}. In this section we obtain the dispersive interaction between two atoms in the framework of Proca QED for any distance regime. For simplicity, we consider the atoms inside a perfectly conducting cube of volume $V$, to be taken to infinity  after performing the calculation. The
electric field operator in Proca QED
is given by\cite{greiner1996field}
\begin{eqnarray}
\mathbf{E}_0(\mathbf{r},t)&=&\sum_{\mathbf{k}\lambda}[\mathbf{E}^{(+)}_{0,\mathbf{k}\lambda}+\mathbf{E}^{(-)}_{0,\mathbf{k}\lambda}]\label{e0}
\end{eqnarray}
with
\begin{equation}
    \mathbf{E}^{(+)}_{0,\mathbf{k}\lambda} = i\sum_{\mathbf{k}\lambda}\left(\frac{2\pi\hbar\omega_k}{V}\right)^{1/2}f_\lambda\boldsymbol{{\epsilon}}_{\mathbf{k}\lambda}a_{\mathbf{k}\lambda}e^{i(\mathbf{r}\cdot\mathbf{k}-\omega_k t)} = \mathbf{E}^{(-)\dagger}_{0,\mathbf{k}\lambda} \, , \label{e0mais}
\end{equation}
where $a_{\mathbf{k}\lambda}$ and $a^{\dagger}_{\mathbf{k}\lambda}$ are
the usual annihilation and creation operators for the mode with wavevector $\boldsymbol{k}$ and polarization $\lambda$, $\omega_k = c\sqrt{k^2+\mu^2}$, $\boldsymbol{\epsilon}_{\mathbf{k}\lambda}$ are the three unitary polarization vectors, with $\lambda = 1,2$ denoting transversal polarizations, while $\lambda=3$ corresponds to the longitudinal one, not present in Maxwell electrodynamics. 
Choosing $\boldsymbol{\epsilon}_{\mathbf{k}\lambda}$  to be normalized requires the introduction of the factor
%
\begin{eqnarray}
f_{\lambda}=\left\{\begin{matrix}
1&& \lambda=1,2\\
\frac{c\mu}{\omega_k}&&\lambda=3 \, ,
\end{matrix}\right.\label{f}
\end{eqnarray}
The interaction between the atoms can be described
 by
the interaction Hamiltonian introduced by P. Milonni \cite{milonni1994quantum}(see specially his Eq. (3.73)), namely
\begin{equation}
H_{int}=-\frac{1}{2}\sum_{\mathbf{k}\lambda}\alpha_A(\omega_k)[\mathbf{E}_{0,\mathbf{k}\lambda}(\mathbf{r}_A)+\mathbf{E}_{B,\mathbf{k}\lambda}(\mathbf{r}_A)]^2 \, , \label{hint3}
\end{equation}
where $\mathbf{E}_{0,\mathbf{k}\lambda}$ is given in Eq.~(\ref{e0})
and $\mathbf{E}_{B,\mathbf{k}\lambda}$ is the electromagnetic field created by
the dipole induced in atom $B$ by the vacuum field, given by (see appendix \ref{dipoleProca})

\begin{eqnarray}
\mathbf{E}_B^{(\pm)}(\mathbf{r}_A,\omega)&=&\frac{e^{\mp ikR}}{R^5}(-k^2R^2\pm3ikR+3)(\mathbf{d}_B^{(\pm)}(\omega_k)\cdot\mathbf{R})\mathbf{R} \, \cr\cr
&+&\frac{e^{\mp ikR}}{R^3}(R^2\omega_k^2\mp ikR-1)\mathbf{d}_B^{(\pm)}(\omega_k), \label{e}
\end{eqnarray}
where the superscript $(\pm)$ follows the same convention as decomposition (\ref{e0}). The dipole operator $\mathbf{d}_B^{(\pm)}(\omega_k)$ inhabits the field Hilbert space and is given by
\begin{equation}
\mathbf{d}_B^{(\pm)}(\omega_k)=\alpha_B(\omega_k)\mathbf{E}^{(\pm)}_{0,\mathbf{k}\lambda}(\mathbf{r}_B). \label{db}
\end{equation}
%


 Physically, Hamiltonian (\ref{hint3}) can be interpreted in the following way: it is the energy of a dipole induced in the atom $A$ due to the field acting in it. This field is the superposition of vacuum electromagnetic field with the electric dipole field generated by atom $B$. This latter dipole is also not permanent but instead induced by the vacuum field according to Eq.(\ref{db}). The interaction energy comes from a first order perturbation calculation, this being the convenience of employing this Hamiltonian. 
%
%
 The dominant order for the interaction comes from the terms involving $\alpha_A\alpha_B$ which are given by
\begin{eqnarray}
H_{int}&=&-\frac{1}{2}\sum_{\mathbf{k}\lambda}\alpha_A(\omega_k)[\mathbf{E}_{0,\mathbf{k}\lambda}(\mathbf{r}_A)\cdot\mathbf{E}_{B,\mathbf{k}\lambda}(\mathbf{r}_A)+\cr\cr
&+&\mathbf{E}_{B,\mathbf{k}\lambda}(\mathbf{r}_A)\cdot\mathbf{E}_{0,\mathbf{k}\lambda}(\mathbf{r}_A)] \, . \label{hint4}\end{eqnarray}
Its expected value yields the interaction energy
\begin{eqnarray}
&U_{\text{total}}\!\!=\!\!-\sum_{\mathbf{k}\lambda}\alpha_A(\omega_k)\alpha_B(\omega_k)g_{ij}(\mathbf{R})\cr\cr
&\times\bra{0} \{E^{i}_{0,\mathbf{k}\lambda}(\mathbf{r}_A),E^{j}_{0,\mathbf{k}\lambda}(\mathbf{r}_B)\} \ket{0},  \label{eint}
\end{eqnarray}
where
\begin{eqnarray}
g_{ij}(\mathbf{R})&=&\text{Re}\,\Bigg\{\frac{e^{-ik R}}{R^3}(R^2\omega^2_k-ik R-1)\delta_{ij}+\cr\cr
&+	&\frac{e^{-ik R}}{R^5}(-k^2R^2+3ik R+3)R_{i}R_{j}\Bigg\} \, . \label{gij}
\end{eqnarray}
From Eqs. (\ref{e0}) and (\ref{e0mais}) we obtain
\begin{eqnarray}
U_{\text{total}}&=&-\text{Re}\sum_{\mathbf{k}\lambda}\left(\frac{2\pi\hbar\omega_k}{V}\right)\alpha_A(\omega_k)\alpha_B(\omega_k)\cr\cr
&&g_{ij}(\mathbf{R})f_\lambda^2(\boldsymbol{\epsilon}_{\mathbf{k}\lambda})_i(\boldsymbol{\epsilon}_{\mathbf{k}\lambda})_j\,e^{i\mathbf{k}\cdot\mathbf{R}}\, .
\end{eqnarray}
%
Substituting Eqs.~(\ref{f}) and (\ref{gij}) into the previous equation, we obtain, after some straightforward algebra,
%
\begin{eqnarray}
U_{\text{total}}&=&-\frac{\hbar}{4\pi^2R^3}\text{Re}\int d^3k\,\alpha_A(\omega_k)\alpha_B(\omega_k)\omega_ke^{-ikR}e^{i\mathbf{k}\cdot\mathbf{R}}\cr\cr
 &\times&\left\{\left[\frac{2\omega_k^2R^2}{c^2}-2ik-2-k^2R^2\sin^2\theta_k\right.\right.\cr\cr
 &+&\left.\left.3ikR\sin^2\theta_k+3\sin^2\theta_k\right]\right.\cr\cr
 &+&\left.\frac{c^2\mu^2}{\omega_k^2}\left[\frac{\omega_k^2R^2}{c^2}-ikR-1-k^2R^2\cos^2\theta_k+\right.\right.\cr\cr
 &+&\left.\left.3ikR\cos^2\theta_k+\cos^2\theta_k\right]\right\}, \label{Utotalpartial}
\end{eqnarray}
where we have defined $\theta_\mathbf{k}$ as the angle between $\mathbf{\hat{k}}$ and $\mathbf{R}$, and took the continuum limit $V\rightarrow\infty$. Defining $x\equiv kR$ and performing the angular integrals, the interaction energy reads
\begin{align}\label{integral}
&U_{\text{total}}=\frac{-\hbar c}{\pi R^7}\int_{0}^{\infty}\,\frac{dx\,x\,\alpha_A\qty(\omega_k(x))\alpha_B\qty(\omega_k(x))}{2\sqrt{x^2+\qty(\mu R)^2}}\times\nonumber\\
&\left\{\left(2 x^4+2 x^2 \left(2 \qty(\mu R)^2-5\right)+3 \left(\qty(\mu R)^4+2\right)\right) \sin (2 x)\right.\nonumber\\
&+\left.4 x \left(x^2-3\right) \cos (2 x)\right\} ,
\end{align}
At this point it is important to make a self-consistent check of our results and reobtain as a particular case the result with usual QED. In fact, by taking the $\mu \to 0$ limit we re-obtain the well known result for the interaction energy within Maxwell QED (see Eq. 3.85 of \cite{milonni1994quantum}).

Integrating by residues in the complex $k$ plane, we may transform the integration in Eq.~\eqref{integral} into an imaginary $k$ one, thus exchanging oscillatory terms by evanescent ones and recasting that equation in a form amenable to numerical integration (see details in appendix \ref{residue}). By expressing our results in terms of the imaginary frequency $\omega = i \frac{c}{R}\zeta$ (with $\zeta\in\mathbb{R}$) we obtain
\begin{eqnarray}\label{imagfreqint}
U_{\text{total}}\!=\!\frac{-\hbar c}{2\pi R^7}\int_{0}^{\infty} d\zeta&\Bigg\{&e^{-2\sqrt{\zeta^2+(\mu R)^2}}f\left(\sqrt{\zeta^2+(\mu R)^2}\right) \times \cr\cr
&\times&\alpha_A\left(i\frac{c}{R}\zeta\right)\alpha_B\left(i\frac{c}{R}\zeta\right)\Bigg\} \, ,   \label{last-complete}
\end{eqnarray}
where
\begin{equation}
f(\xi)\!=\!6+3\qty(\mu R)^4+12\xi+(10\!-\!4\qty(\mu R)^2)\xi^2+4\xi^3+2\xi^4.
\end{equation}
The mass of the photon has a crucial role in the calculation, since it introduces a cut due to the square root in the dispersion relation $\omega_k = c\sqrt{k^2+\mu^2}$. 

\subsection{Qualitative discussion\label{qualitative}}

A striking feature of Eq.~\eqref{imagfreqint} concerns the relevant physical scales. Pure dimensional analysis shows the existence of three timescales, corresponding to three typical frequencies: $c/R$, $c\mu$, and the dominant atomic transition frequency $\omega_{0}$. Nonetheless, an inspection of the integrand of Eq.~\eqref{imagfreqint} reveals that for the dispersive interaction only two scales are involved: $\omega_0$, which is present in the polarizability functions and  
\begin{equation}
    \omega_{F} \sim c\sqrt{\frac{1}{R^2}+\mu^2}\, .
    \label{omegafield}
\end{equation}
which is contained in the field functions $f$ and the decaying exponential. Let us now employ these scales in order to determine the values of $\zeta$ which dominate the integral in Eq.~\eqref{last-complete}. 
Equation~\eqref{kramers-heisenberg} shows that the atoms become transparent ($\alpha_{A,B}\approx 0$) at large imaginary frequencies, $(c/R)\,\zeta\gg \omega_0$, suppressing large values of $\zeta$.  
 The field modes are also suppressed for high frequencies - this is the main advantadge of the rotation in the complex plane discussed in Appendix \ref{residue}. This yields to an exponential attenuation, cutting off $(c/R)\,\zeta \gg \omega_{F}$ in Eq.~\eqref{last-complete}. These frequencies correspond to large wavevectors $k\gg R^{-1}$ which oscillate a large number of times between atoms $A$ and $B$, leading to a decorrelation between the field at positions $\bm r_A$ and  $\bm r_B$ after superposing a continuum of modes. The interplay between large-frequency transparency and large-wavenumber decorrelation leads to two opposite asymptotic regimes, which we discuss below.

These frequencies correspond to large wavevectors $k\gg R^{-1}$ which oscillate a large number of times between atoms $A$ and $B$, which, upon superposing a continuum of modes, leads to decorrelation between the field at position $\bm r_A$ and the atomic dipole at position $\bm r_B$. 
The interplay between large-frequency transparency and large-wavenumber decorrelation leads to two opposite asymptotic regimes, which we discuss below.

When $\omega_{0}\gg \omega_{F}$, the contribution of the high frequency in integrand ~(\ref{imagfreqint}) is suppressed by decorrelation in the response of the field before atomic high-frequency transparency sets in. Therefore, interatomic dipole-dipole correlations are limited by the retardation of the electromagnetic field.
Moreover, from Eq.~(\ref{kramers-heisenberg}), one may approximate $\alpha(\omega) \approx \alpha(0)$, where the latter denotes the static polarizability. Physically this is due to the fact that in this regime the atom easily follows the much slower field oscillations. In other words, in this regime the correlation between the atomic dipoles are limited by the delay in electromagnetic interaction. This case is denominated the retarded limit, known as the Casimir-Polder regime in the case of Maxwell electrodynamics. Note that in Proca QED this regime is possible only for small masses $\mu \ll \omega_0/c$.    %

In the opposite limit, $\omega_{0}\ll \omega_{F}$, large values of $\zeta$ are suppressed due to high-frequency transparency before field-decorrelation or retardation effects become relevant. We can therefore  approximate the imaginary frequency to be $\zeta\approx 0$ for the field --- that is, everywhere in the integrand of Eq.(\ref{last-complete}) except in the polarizabilities. 
In this case we may treat the electromagnetic field in the electrostatic approximation, thus recovering the results of section \ref{Non-retarded regime}, as we shall explicitly confirm in section\ref{subsec:non-retarded limit}. From Eq.~(\ref{omegafield}) we see that the non-retarded regime applies to a wider range of frequencies in Proca than in Maxwell electrodynamics, since the condition for $R$ is weaker. In particular, if $\mu c\gg \omega_0$  we are always in the non-retarded limit, regardless of distance. Naturally, however, for $R\gg \mu^{-1} $ the interaction is exponentially suppressed and thus very weak.   %
 
By defining the phase velocity $v_p(k) \equiv \omega(k)/k = c \sqrt{(1 + \mu^2/k^2)}$, the non-retarded regime can be written as $\omega_0 R\ll v(k_{\rm field})$ where $k_{\rm field}=1/R$ (with an analogous expression for the retarded regime). In this way, the condition defining each regime in Proca QED is the same as in the Maxwellian case as long as in the latter we substitute $c$ by the phase velocity of the photon evaluated at the wavenumber $1/R$. Nonetheless there are now remarkable differences. In Maxwell, $t_{\gamma} = R/c$ is the time it takes to a photon to travel the distance separating the atoms and corresponds to the retardation time of the interaction. Therefore, the non-retarded condition means that $t_{\gamma}$ must be much smaller than the atomic typical time scale $1/\omega_0$, representing the dipole's fluctuation time. No such simple interpretation holds in Proca, since due to dispersion there is not a single retardation time describing the interaction and the phase velocity can not be interpreted as the velocity of propagation of the electromagnetic wave. Indeed, phase velocity is superluminal in Proca electrodynamics. This poses no paradox since we are dealing only with virtual photons and not with any information propagation. The key concept here is that the photon mass increases the photon frequency for every wavelength and thus retardation effects are less relevant in Proca electrodynamics than in the Maxwellian case, since it is the slower dynamics which is more relevant to limit the dipole-dipole correlation. We now turn to the detailed calculation which supports the above statements.

\subsection{Retarded regime ($\omega_f\ll \omega_0$)}\label{Subsec: Retarded limit}

In this regime, we may substitute the atomic polarizabilities given in Eq.~(\ref{kramers-heisenberg}) by their static values $\alpha(\omega=0)$. With this approximation, the integral \eqref{last-complete} takes the form
\begin{eqnarray}
U_{\text{Ret}} &=& \frac{-\hbar c}{2\pi R^7}\alpha_A(0)\alpha_B(0) \cr\cr
&\times&\int_{0}^{\infty} d\zeta\, e^{-2\sqrt{\zeta^2+(\mu R)^2}}f\left(\sqrt{\zeta^2+(\mu R)^2}\right) \, .
\end{eqnarray}
After performing the integration we are left with
%
\begin{eqnarray}\label{retarded-energy}
&U_{\text{Ret}}&=-\frac{\hbar c \alpha _A \alpha _B }{4 \pi  R^7}\left\{ 8 \qty(\mu R)^2 \left(\qty(\mu R)^2+3\right) K_0(2\mu R)+\right.\cr\cr
&+&\left(2 \qty(\mu R)^5+31 \qty(\mu R)^3+24(\mu R)\right) K_1(2 \mu R)+\cr\cr
&+&\left.22 \qty(\mu R)^2 K_2(2 \mu R)\right\},
\end{eqnarray}
where $K_{\nu}(x)$ are the modified Bessel functions of the second kind.

 As $\mu R\rightarrow 0$, the term $(\mu R)K_0(2\mu R)$ in Eq.~\eqref{retarded-energy} becomes subdominant, while, using  $K_{\alpha} (x) \sim \frac{\Gamma(
\alpha)}{2}\left(\frac{2}{x}\right)^{\alpha}$, valid for $\alpha > 0$, we see that the term inside the brackets becomes $\sim (24+22)/2=23$, thus re-obtaining the famous result by  Casimir and Polder \cite{casimir1948influence} hereafter denoted $U_{\rm CP}$. Working out the next term in the small mass limit $\mu R \ll 1$ we obtain
\begin{equation}\label{eq:asymp-r}
U=U_{\rm CP}\left(1-\frac{15\qty(\mu R)^2}{46}\right)+\mathcal{O}[\qty(\mu R)^4].
\end{equation}
\begin{figure}[h!]
    \centering
    \includegraphics[width=8.6cm]{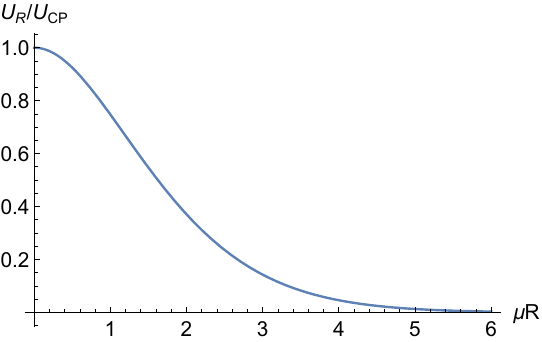}
    \caption{Interaction potential normalized by the Casimir-Polder potential $U_{\text{CP}}$ as a function of the dimensionless radius parameter $\mu R$.}
    \label{plot-rUxR}
\end{figure}
\begin{figure}
    \centering
    \includegraphics[width=8.6cm]{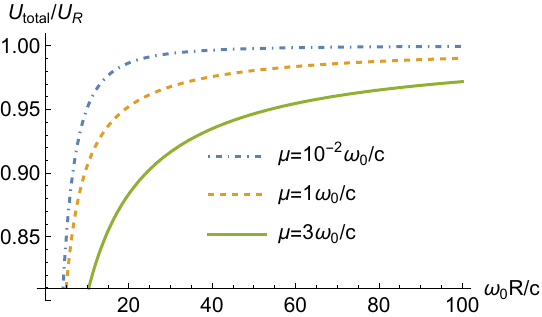}
    \caption{Complete interaction energy $U_{\text{total}}$, normalized by the retarded interaction $U_R$, as a function of $\omega_0 R/c$, for identical two-level atoms with transition frequency $\omega_0 R$ for different values of the photon mass.}
    \label{numerical retarded}
\end{figure}
%


%

As we had obtained for the non-retarded regime, the first correction is of order $(\mu R)^2$ and weakens the interaction. This is due to the fact that the mass increases the frequency for every wavevector, thus weakening the correlation between the atoms. This behavior is illustrated in Fig. \ref{plot-rUxR} where we plot the Casimir-Polder retarded energy normalized by the Casimir-Polder result in terms of $\mu R$.

In the opposite limit, $\mu R \gg 1$, we have an exponential decay given by the asymptotic form of the modified Bessel functions,
\begin{equation}
    U =-\frac{\hbar c \alpha _A(0) \alpha _B(0)\mu^{9/2}}{4 \pi ^{1/2} R^{5/2}} e^{-2\mu R} \left\{ 1+\mathcal{O}\left(\frac{1}{\mu R}\right)\right\} \label{fracpot}
\end{equation}
The previous result shows that, for a given distance between the atoms, their interaction potential  decreases exponentially with the mass of the photon, as expected, since the quantum fluctuations of the field are drastically suppressed as the photon mass increases. Note also the fractional power law dependence with the distance between the atoms. This fact resembles somehow the correction to the Coulomb interaction which appears in  the so called Uehling potential (see \cite{peskin} for more detail). In the last case, the electron loop in the vacuum polarization Feynman diagram introduces the electron mass scale ($m_e$) into the problem and leads, for large $R$  ($R\gg 1/m_e$), to a correction to the Coulomb potential which is suppressed expeonentially with $m_eR$ and which falls with the distance with the same fractional power law that we have obtained for a massive photon. 

In Fig. \ref{numerical retarded} we plot the exact interaction energy given by Eq.(\ref{last-complete}) normalized by $U_{\rm Ret}$ written in Eq.(\ref{retarded-energy}) for identical two level atoms as a function of $\omega_0R$, where $\omega_0$ denotes the atomic transition frequency. This is done by assuming that only a single term is present in the polarizability defined in Eq.(\ref{kramers-heisenberg}). Notice that the Casimir-Polder approximation overestimates the interaction energy, which is expected since the static polarizability is higher than the dynamical one far from resonance (which is always the case in the retarded regime). Another striking feature is that the higher the transition frequency is, for a fixed distance, the worse the retarded approximation becomes, along the lines discussed in the previous subsection.

\subsection{Non-retarded regime ($\omega_f\gg \omega_0$)\label{subsec:non-retarded limit}}

Here we take the opposite limit and, as discussed in section \ref{qualitative}, we take the zero frequency limit for the field degrees of freedom, setting $\zeta=0$ in Eq.~(\ref{last-complete}), thus obtaining for interaction energy in the non-retarded regime the expression
\begin{eqnarray}
U_{\text{NR}} (R)\!\!=\!\!\frac{-\hbar c e^{-2\mu R}}{2\pi R^7}p(\mu R)I\, , \label{NRfromcomplete}
\end{eqnarray}
where $p$ is defined in Eq.~(\ref{P}) and
\begin{eqnarray}
    I &=& \int_{0}^{\infty} d\zeta\, \alpha_A\left(\frac{i\zeta c}{R}\right)\alpha_B\left(\frac{i\zeta c}{R}\right)\cr\cr
    &=&\frac{4}{9\hbar^2}\sum_{r,s\neq 0}|\boldsymbol{d}_{A}^{0r}|^2|\boldsymbol{d}_{B}^{0r}|^2 \int_0^\infty \frac{d\zeta\, \omega_{r0}\omega_{s0}}{\left(\omega_{r0}^2+\frac{R^2\zeta^2}{c^2}\right)\left(\omega_{s0}^2+\frac{R^2\zeta^2}{c^2}\right)} \cr\cr
    &=& \frac{2\pi R}{9\hbar c}\Lambda \, ,
    \label{eq:nonretint}
\end{eqnarray}
where $\Lambda$ is defined in Eq.~(\ref{Lambda}). 

Substituting the result in Eq.~\eqref{eq:nonretint} into Eq.~(\ref{NRfromcomplete}) we reobtain Eq.~(\ref{nonretarded}) of section \ref{Non-retarded regime}. Therefore, in Proca we see that when the retardation of the electromagnetic field can be neglected we do not need to quantize the electromagnetic field, as it is also true in Maxwell. Nonetheless, in Proca we have an extra subtlety worth mentioning. In section \ref{Non-retarded regime} we did not quantize the classical degrees of freedom for the field, so that the quantum fluctuations came
entirely from the atomic dipole fluctuations. In particular, since we are in the electrostatic regime ($\omega=0$), the field produced by the fluctuating dipoles is evanescent. In section \ref{fullcalculation} on the other hand, our Hamiltonian (\ref{hint4}) contained only field fluctuations with the atomic dipoles being induced by them, as in Eq.~(\ref{db}). Here we have only propagating modes for the fields, which requires frequencies greater than $\mu c$, in marked contrast with the previous calculation. Nevertheless, as we have demonstrated in this section, the results of the electrostatic treatment of section \ref{Non-retarded regime} are contained in the full quantum electrodynamics analysis. This is physically expected since we can choose whether we consider the interatomic interaction as fluctuating dipoles which generates electric field or the other way around, with the vacuum electromagnetic fluctuations playing the leading role. We could as well start from a democratic Hamiltonian which involved fluctuations of both degrees of freedom but then we would have to perform a tedious 4$^{\rm th}$ order perturbation theory calculation \cite{mattioli2019casimir}. 
%
\begin{figure}
    \centering
    \includegraphics[width=8.6cm]{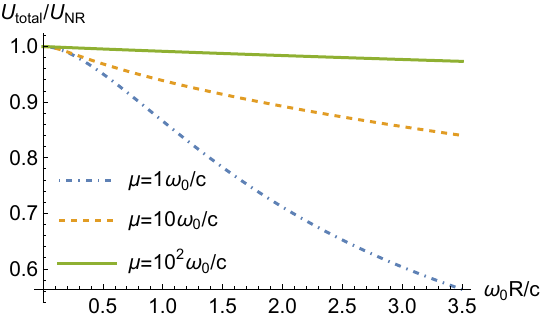}
     \caption{Complete interaction energy $U_{\text{total}}$, normalized by the non-retarded regime interaction $U_{NR}$, as a function of $\omega_0 R/c$ for different values of the photon mass. In this analysis, atoms $A$ and $B$ are considered to be identical two-level systems of energy gap $\omega_0$. }
    \label{numerical non-retarded}
\end{figure}

The ratio between the full result from Eq.~\eqref{last-complete} for identical two-level atoms and the interaction energy in the non-retarded approximation  is shown as a function of $\mu R$ in Fig.~\ref{numerical non-retarded}. We note that, as $\mu c/\omega_0$ increases, the non-retarded regime extends to larger values of $\omega_0 R/c$, but the interaction energy itself becomes more strongly suppressed for $R \gtrsim c/\omega_0$. 

\begin{figure}
    \centering
    \includegraphics[width=8.6cm]{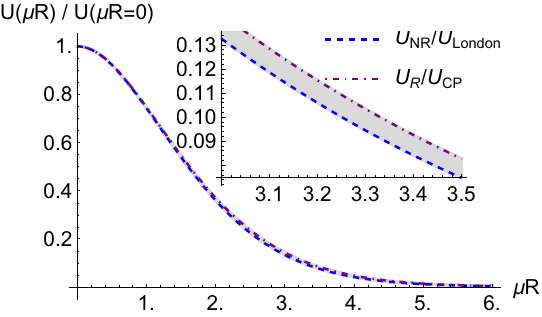}
     \caption{Ratio between the interaction energies for massive and massless QED, as a function of the dimensionless radius parameter $\mu R$ for two regimes: the non-retarded (blue dashed line) and  the retarded (purple dot-dashed line) approximations. The gray band visible on the inset displays the curves $U_{\rm total}(\mu R)/U_{\rm total}(\mu R =0)$ for $\omega_0$ in the range $10^{-4}c/R-10^4c/R$.}
    \label{band}
\end{figure}

The attentive reader may have noticed the similarity between Figs.  \ref{plot-nr} and \ref{plot-rUxR}, which show the ratio between the massive and massless interaction energies, $U(\mu R)/U(\mu R =0)$, evaluated in the non-retarded and the retarded approximations, respectively. As shown in Fig.~\ref{band}, the curves for this ratio as a function of the dimensionless parameter $\mu R$ are indeed remarkably similar for both regimes, even though the two approximations correspond to opposite limits. In the particular case where $\mu R\ll 1$, this can be readily verified by comparing Eqs.(~\ref{eq:asymp-nr}) and \eqref{eq:asymp-r}, since the coefficients of the quadratic correction in $\mu R$ differ by less than 1\% 
\footnote{We note that the asymptotic expressions for $\mu R\gg 1$ have different behaviors for the two approximations. However, this difference only becomes pronounced for small values of $U(\mu R)/U(\mu R =0)$, which explains why they are not visible in Fig.~\ref{band}.}.
This happens because for a given interatomic distance, the ratio $U_{\rm total}(\mu R)/U_{\rm total}(\mu R =0)$, evaluated from Eq.~(\ref{last-complete})  is very nearly independent of the transition frequency $\omega_0$. Indeed, by varying $\omega_0$ between $10^{-4}c/R$ and $10^4c/R$ for every distance we obtain the grey band depicted in Fig.~\ref{band} connecting the retarded curve (which is the limite for $\omega_0\rightarrow \infty)$ with the non-retarded curve (corresponding to the $\omega_0\rightarrow 0$ limit).

\section{Conclusions and final remarks}\label{sec:resultsandfinalremarks}

In this paper we have derived the dispersive interaction energy between two polarizable atoms, with no permanent dipole moments, within Proca electrodynamics for any distance regime. We emphasize that we employed a very convenient Hamiltonian, introduced in Ref.\cite{milonni1994quantum}, which allows for a first order perturbative calculation. 

%
We have found that, despite the inclusion of an extra energy scale corresponding to the photon mass, dispersive interactions within Proca electrodynamics are still marked by the existence of no more than two asymptotic regimes, corresponding to the London and Casimir-Polder limits of Maxwellian electrodynamics.
A timescale consisting of interatomic distance over phase velocity may be used to distinguish between these two regimes, but, in contrast to Maxwell's theory,  the nonlinear relationship between frequency and wavenumber in the massive case prevents its unambiguous physical interpretation as a retardation time. This is due to the fact that the phase velocity is both frequency-dependent and  superluminal in the massive theory.
We have computed closed-form analytical expressions that approximate the dispersive interaction energy in the retarded and non-retarded regimes and presented a detailed discussion on their differences with respect to the case of a massless photon.
Our results reveal an enhancement of the non-retarded regime, which extends over a broader range of interatomic distances in the presence of a photon mass.

 We have also shown that the effects of the photon mass weaken the interaction energy. More strikingly, we have shown that, for a given interatomic distance, this weakening is very similar for both the retarded and the non-retarded regimes. This means that the mass correction is nearly independent of the atomic transition frequency. As a perspective, it would be interesting to explore if this still holds beyond the atomic scenario, analyzing for instance the Casimir-Proca force between dielectric media.

Our results should be relevant not only to beyond the standard model physics scenarios, but also for any theories or effective descriptions featuring massive vector bosons or an effective photon mass.
An effective photon mass may emerge, for instance, inside electrolyte solutions, metals or waveguides \cite{chiao1999bogoliubov,Souza2015}.
Other massive vector bosons can be found in effective descriptions of nuclear matter, such as relativistic mean-field models of nuclear interactions
 \cite{Walecka:1974qa,Boguta:1982wr,norman1997compact,Horowitz:2000xj,Chen:2014sca}.
In the nuclear context, the possible polarization of mesons into a baryon-numer dipole could lead to new couplings with vector mesons, with potentially interesting phenomenological consequences \cite{Pisarski:2021aoz}.

We hope our findings will inform future investigations on potential observable effects of a finite, albeit small, photon mass \cite{mattioli2019casimir}.
Because bounds on the mass of the photon are rather small, the detection of a photon mass via van der Waals interactions would be far from trivial. Nonetheless, possible implications of the finite-mass modifications in the condensation of gases and other physical phenomena could potentially prove more revealing of the photon mass then the van der Waals interaction energy in itself. It would be interesting to explore this possibility in future work.

\begin{acknowledgments}
 The authors would like to thank Dr. Pedro C. Malta for the careful reading of this manuscript. This study was financed in part by the Coordena\c{c}\~{a}o de Aperfei\c{c}oamento de Pessoal de N\'{\i}vel Superior - Brasil (CAPES) - Finance Code 001. C.A.D.Z. are partially supported by Conselho Nacional de Desenvolvimento Cient\'{\i}fico e Tecnol\'{o}gico (CNPq) under the grant   310703/2021-2.  C.A.D.Z. is also partially supporteprocessa\c{c}\~{a}o Carlos Chagas Filho de Amparo \`{a} Pesquisa do Estado do Rio de Janeiro (Faperj) under Grant E-26/201{.}447/2021 (Programa Jovem Cientista do Nosso Estado). 
M.H. was supported in part by the National Science Foundation (NSF) within the framework of the MUSES Collaboration, under grant number OAC-2103680, and by Universidade Estadual do rio de Janeiro, within the Programa de Apoio à Docência (PAPD).  
\end{acknowledgments}

\appendix

\section{The Proca electric dipole field\label{dipoleProca}}


In this appendix we obtain the Proca field generated by an electric dipole, following an alternative approach to the one presented in Ref.\cite{dragulin2008green}. We work in the Fourier frequency domain, but keep the dependence in spatial variables. With this choice  the potential quadrivector $(\phi,\boldsymbol{A})$ satisfies
in Proca electrodynamics the equation
\begin{equation}
\left(\boldsymbol{\nabla}^2+\frac{\omega^2}{c^2}-\mu^2\right)A^{\mu}(\boldsymbol{r})=-\frac{4\pi}{c}j^{\mu}(\boldsymbol{r}) \, , \label{boxAmu}
\end{equation}
which can be readily solved with the aid of the Green function satisfying
\begin{equation}
    \left(\boldsymbol{\nabla}^2+\frac{\omega^2}{c^2}-\mu^2\right)G(\boldsymbol{r},\boldsymbol{r}',\omega)=-4\pi \delta(\boldsymbol{r}-\boldsymbol{r}') \, ,
\end{equation}
whose solution is given by
\begin{equation}
G(\mathbf{r}, \mathbf{r'},\omega)=\frac{e^{\mp ik|\mathbf{r}-\mathbf{r'}|}}{|\mathbf{r}-\mathbf{r'}|} \label{g} \, ,
\end{equation}
where we have definied $k:=\sqrt{\frac{\omega^2}{c^2}-\mu^2}$ and
the minus (plus) sign refers to the solution with (positive) negative frequency. For the purposes of this paper, we need only propagating frequencies $\omega > \mu c$ for these are the only ones present in the vacuum electromagnetic field. The Proca field produced by a dipole follows from solving Eq.~(\ref{boxAmu}) with
\begin{equation}
    j^{\mu}=(-c\boldsymbol{d}(\omega)\cdot\boldsymbol{\nabla}\delta(\boldsymbol{r}-\boldsymbol{r}') , -i\omega\boldsymbol{d}(\omega)\delta(\boldsymbol{r}-\boldsymbol{r}') )\, ,
\end{equation}
which corresponds to a current 4-vector of a point dipole\cite{Pitombo2021} $\boldsymbol{d}(\omega)$ at position $\boldsymbol{r}'$. Hence, the potentials are given by
\begin{eqnarray}
    \phi (\boldsymbol{r},\omega) &=& \boldsymbol{d}(\omega)\cdot\boldsymbol{\nabla}'G(\boldsymbol{r},\boldsymbol{r}',\omega) \, , \label{phi} \\
     \boldsymbol{A} (\boldsymbol{r},\omega) &=& -\frac{i\omega}{c}\boldsymbol{d}(\omega)G(\boldsymbol{r},\boldsymbol{r}',\omega) \, . \label{A}
\end{eqnarray}
Hence, the Proca electric field for a dipole is given by
\begin{equation}
E_i(\mathbf{r}, \omega)=d_j(\omega)\left(\partial_i\partial_j+\delta_{ij}\omega^2\right)G(\mathbf{r}, \mathbf{r'},\omega) \label{eif} \, ,
\end{equation}
where we have employed $\partial_j'=-\partial_j$, which is valid since $G$ - given in Eq.~(\ref{g}) - is a function only of $\boldsymbol{r}-\boldsymbol{r}'$. By evaluating the derivatives present in the previous equation we obtain Eq.~(\ref{e}) of the main paper. Finally, the electrostatic case is obtained by taking the static limit $\omega\rightarrow 0$, which is equivalent to take $k = \mp \mu$. The choice of the sign is made to comply with the boundary condition $G\rightarrow 0$ for $r\rightarrow\infty$. With this choice, Eq.~(\ref{e}) furnishes Eq.~(\ref{estatico}) of section \ref{Non-retarded regime}.

\section{The integral of Sec. \ref{fullcalculation} \label{residue}}

In this appendix we will explicitly calculate integrals of the form
\begin{equation}\label{intB}
I=A\int_0^{\infty} \frac{dx}{\sqrt{x^2+y^2}}\left[P_s(x)\sin(2x)+P_c(x)\cos(2x)\right]  ,
\end{equation}
where $A$ is some real constant, $P_s(x)$ is a real odd function and $P_c(x)$ is a real even function with no branch cuts. Those integrals are clearly divergent on the UV regime, so we will need to perform it on the extended complex plane. 

The first step is simply to use the parity of the integrand to write the above integral as
\begin{equation}\label{intB2}
I=\frac{A}{2}\text{Re}\int_{-\infty}^{\infty} \frac{dx}{\sqrt{x^2+y^2}} \left[P_c(x)-iP_s(x)\right]e^{2ix}. 
\end{equation}

We then move to the complex plane, defining a complex variable $z$ such that $\Re z = x$, so that integral \eqref{intB2} lies on the real $z$ line. 
\begin{figure}[h!]
    \centering
    \includegraphics[width=8.6cm]{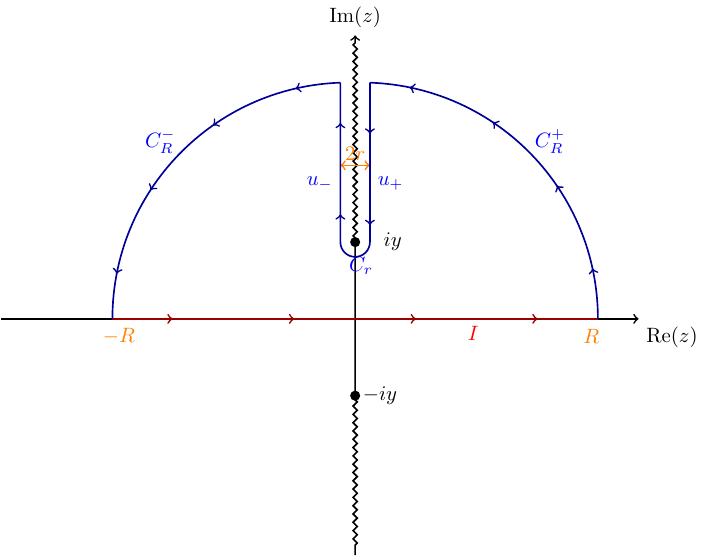}
    \caption{Contour in the complex plane used to compute Eq. \eqref{intB2}.}
    \label{appendixBfigure}
\end{figure}
The above integrand has two branch cuts on the complex plane (one running from $x=iy$ to $x=i\infty$ and the other running from $x=-iy$ to $x=-\infty$ ), as shown in Fig.~\ref{appendixBfigure}. To apply the residue theorem, we avoid these branch cuts, by employing the contour shown in Fig.~\ref{appendixBfigure}. We choose to close the contour on the upper half plane, so that the segments $C_R^{\pm}$ on this figure evaluate to zero at complex infinity.
%
The only parts of the contour that are non-zero are $I$, $u^{+}$ and $u^{-}$, and the residue theorem implies that
\begin{align}
&\int_I g(z)\,dz+\int_{u_+}g(z)\,dz+\int_{u_-}g(z)\,dz=0    ,
\end{align}
where
\begin{equation}
g(z)=\frac{\tilde{P}(z)\,e^{2iz}}{\sqrt{z^2+y^2}}
\end{equation}
and
\begin{equation}
\tilde{P}(z)=P_c(z)-iP_s(z) .
\end{equation}

The branch cut is chosen such that the square root has positive imaginary part on the right side  (the $u^{-}$ line) and negative imaginary part on the left side (the $u^{-}$ line). Hence,
\begin{align}
&\int_{u_+}g(z)dz=\frac{A}{2}\int_{\infty}^{y}\frac{(idu)}{i\sqrt{u^2-y^2}}\tilde{P}(iu)e^{-2u}\nonumber\\
&= -\frac{A}{2}\int_{y}^{{\infty}}\frac{du}{\sqrt{u^2-y^2}}\tilde{P}(iu)\,e^{-2u}
\end{align}
and
\begin{align}
&\int_{u_-}g(z)dz=\frac{A}{2}\int_{y}^{\infty}\frac{(idu)}{-i\sqrt{u^2-y^2}}\tilde{P}(iu)e^{-2u}\nonumber\\
&= -\frac{A}{2}\int_{y}^{{\infty}}\frac{du}{\sqrt{u^2-y^2}}\tilde{P}(iu)e^{-2u} \\
&= \int_{u_+}g(z)dz
\end{align}
So that integral $I$ becomes:
\begin{equation}\label{intB3}
I=-A\,\text{Re}\int_{y}^{{\infty}}\frac{du}{\sqrt{u^2-y^2}}\tilde{P}(iu)\,e^{-2u}.
\end{equation}



\begin{thebibliography}{60}%
\makeatletter
\providecommand \@ifxundefined [1]{%
 \@ifx{#1\undefined}
}%
\providecommand \@ifnum [1]{%
 \ifnum #1\expandafter \@firstoftwo
 \else \expandafter \@secondoftwo
 \fi
}%
\providecommand \@ifx [1]{%
 \ifx #1\expandafter \@firstoftwo
 \else \expandafter \@secondoftwo
 \fi
}%
\providecommand \natexlab [1]{#1}%
\providecommand \enquote  [1]{``#1''}%
\providecommand \bibnamefont  [1]{#1}%
\providecommand \bibfnamefont [1]{#1}%
\providecommand \citenamefont [1]{#1}%
\providecommand \href@noop [0]{\@secondoftwo}%
\providecommand \href [0]{\begingroup \@sanitize@url \@href}%
\providecommand \@href[1]{\@@startlink{#1}\@@href}%
\providecommand \@@href[1]{\endgroup#1\@@endlink}%
\providecommand \@sanitize@url [0]{\catcode `\\12\catcode `\$12\catcode `\&12\catcode `\#12\catcode `\^12\catcode `\_12\catcode `\%12\relax}%
\providecommand \@@startlink[1]{}%
\providecommand \@@endlink[0]{}%
\providecommand \url  [0]{\begingroup\@sanitize@url \@url }%
\providecommand \@url [1]{\endgroup\@href {#1}{\urlprefix }}%
\providecommand \urlprefix  [0]{URL }%
\providecommand \Eprint [0]{\href }%
\providecommand \doibase [0]{https://doi.org/}%
\providecommand \selectlanguage [0]{\@gobble}%
\providecommand \bibinfo  [0]{\@secondoftwo}%
\providecommand \bibfield  [0]{\@secondoftwo}%
\providecommand \translation [1]{[#1]}%
\providecommand \BibitemOpen [0]{}%
\providecommand \bibitemStop [0]{}%
\providecommand \bibitemNoStop [0]{.\EOS\space}%
\providecommand \EOS [0]{\spacefactor3000\relax}%
\providecommand \BibitemShut  [1]{\csname bibitem#1\endcsname}%
\let\auto@bib@innerbib\@empty
\bibitem [{\citenamefont {Milonni}(1994)}]{milonni1994quantum}%
  \BibitemOpen
  \bibfield  {author} {\bibinfo {author} {\bibfnamefont {P.~W.}\ \bibnamefont {Milonni}},\ }\href@noop {} {\emph {\bibinfo {title} {The Quantum Vacuum, an introduction to quantum electrodynamics. Academic Press Inc.}}}\ (\bibinfo  {publisher} {Academic Press},\ \bibinfo {year} {1994})\BibitemShut {NoStop}%
\bibitem [{\citenamefont {Israelachvili}(2011)}]{israelachvili}%
  \BibitemOpen
  \bibfield  {author} {\bibinfo {author} {\bibfnamefont {J.~N.}\ \bibnamefont {Israelachvili}},\ }\href@noop {} {\emph {\bibinfo {title} {Intermolecular and Surface Forces,}}}\ (\bibinfo  {publisher} {Academic Press},\ \bibinfo {year} {2011})\BibitemShut {NoStop}%
\bibitem [{\citenamefont {Buhmann}(2013{\natexlab{a}})}]{buhmann1}%
  \BibitemOpen
  \bibfield  {author} {\bibinfo {author} {\bibfnamefont {S.~Y.}\ \bibnamefont {Buhmann}},\ }\href@noop {} {\emph {\bibinfo {title} {Dispersion Forces I: Macroscopic quantum electrodynamics and ground-state Casimir, Casimir--Polder and van der Waals forces}}},\ Vol.\ \bibinfo {volume} {247}\ (\bibinfo  {publisher} {Springer},\ \bibinfo {year} {2013})\BibitemShut {NoStop}%
\bibitem [{\citenamefont {Buhmann}(2013{\natexlab{b}})}]{buhmann2}%
  \BibitemOpen
  \bibfield  {author} {\bibinfo {author} {\bibfnamefont {S.~Y.}\ \bibnamefont {Buhmann}},\ }\href@noop {} {\emph {\bibinfo {title} {Dispersion Forces II: Many-Body Effects, Excited Atoms, Finite Temperature and Quantum Friction}}},\ Vol.\ \bibinfo {volume} {248}\ (\bibinfo  {publisher} {Springer},\ \bibinfo {year} {2013})\BibitemShut {NoStop}%
\bibitem [{\citenamefont {Milton}(2001)}]{milton}%
  \BibitemOpen
  \bibfield  {author} {\bibinfo {author} {\bibfnamefont {K.~A.}\ \bibnamefont {Milton}},\ }\href@noop {} {\emph {\bibinfo {title} {The Casimir Effect: Physical Manifestations of Zero-Point Energy}}}\ (\bibinfo  {publisher} {World Scientific},\ \bibinfo {year} {2001})\BibitemShut {NoStop}%
\bibitem [{\citenamefont {Geim}\ and\ \citenamefont {Grigorieva}(2013)}]{Geim2013}%
  \BibitemOpen
  \bibfield  {author} {\bibinfo {author} {\bibfnamefont {A.~K.}\ \bibnamefont {Geim}}\ and\ \bibinfo {author} {\bibfnamefont {I.~V.}\ \bibnamefont {Grigorieva}},\ }\href {https://doi.org/10.1038/nature12385} {\bibfield  {journal} {\bibinfo  {journal} {Nature}\ }\textbf {\bibinfo {volume} {499}},\ \bibinfo {pages} {419} (\bibinfo {year} {2013})}\BibitemShut {NoStop}%
\bibitem [{\citenamefont {Castillo-L{\'o}pez}\ \emph {et~al.}(2024)\citenamefont {Castillo-L{\'o}pez}, \citenamefont {Esquivel-Sirvent}, \citenamefont {Pirruccio},\ and\ \citenamefont {Villarreal}}]{castillo2024casimir}%
  \BibitemOpen
  \bibfield  {author} {\bibinfo {author} {\bibfnamefont {S.~G.}\ \bibnamefont {Castillo-L{\'o}pez}}, \bibinfo {author} {\bibfnamefont {R.}~\bibnamefont {Esquivel-Sirvent}}, \bibinfo {author} {\bibfnamefont {G.}~\bibnamefont {Pirruccio}},\ and\ \bibinfo {author} {\bibfnamefont {C.}~\bibnamefont {Villarreal}},\ }\href {https://doi.org/10.3390/physics6010026} {\bibfield  {journal} {\bibinfo  {journal} {Physics}\ }\textbf {\bibinfo {volume} {6}},\ \bibinfo {pages} {394} (\bibinfo {year} {2024})}\BibitemShut {NoStop}%
\bibitem [{\citenamefont {Castillo-L{\'o}pez}\ \emph {et~al.}(2022)\citenamefont {Castillo-L{\'o}pez}, \citenamefont {Esquivel-Sirvent}, \citenamefont {Pirruccio},\ and\ \citenamefont {Villarreal}}]{castillo2022casimir}%
  \BibitemOpen
  \bibfield  {author} {\bibinfo {author} {\bibfnamefont {S.~G.}\ \bibnamefont {Castillo-L{\'o}pez}}, \bibinfo {author} {\bibfnamefont {R.}~\bibnamefont {Esquivel-Sirvent}}, \bibinfo {author} {\bibfnamefont {G.}~\bibnamefont {Pirruccio}},\ and\ \bibinfo {author} {\bibfnamefont {C.}~\bibnamefont {Villarreal}},\ }\href {https://doi.org/10.1038/s41598-022-06866-5} {\bibfield  {journal} {\bibinfo  {journal} {Scientific Reports}\ }\textbf {\bibinfo {volume} {12}},\ \bibinfo {pages} {2905} (\bibinfo {year} {2022})}\BibitemShut {NoStop}%
\bibitem [{\citenamefont {Fosco}\ \emph {et~al.}(2024)\citenamefont {Fosco}, \citenamefont {Lombardo},\ and\ \citenamefont {Mazzitelli}}]{fosco2024casimir}%
  \BibitemOpen
  \bibfield  {author} {\bibinfo {author} {\bibfnamefont {C.~D.}\ \bibnamefont {Fosco}}, \bibinfo {author} {\bibfnamefont {F.~C.}\ \bibnamefont {Lombardo}},\ and\ \bibinfo {author} {\bibfnamefont {F.~D.}\ \bibnamefont {Mazzitelli}},\ }\href {https://doi.org/10.3390/physics6010020} {\bibfield  {journal} {\bibinfo  {journal} {Physics}\ }\textbf {\bibinfo {volume} {6}},\ \bibinfo {pages} {290} (\bibinfo {year} {2024})}\BibitemShut {NoStop}%
\bibitem [{\citenamefont {Fosco}\ and\ \citenamefont {Mazzitelli}(2020)}]{fosco2020casimir}%
  \BibitemOpen
  \bibfield  {author} {\bibinfo {author} {\bibfnamefont {C.~D.}\ \bibnamefont {Fosco}}\ and\ \bibinfo {author} {\bibfnamefont {F.~D.}\ \bibnamefont {Mazzitelli}},\ }\href {https://doi.org/10.1103/PhysRevD.101.045012} {\bibfield  {journal} {\bibinfo  {journal} {Physical Review D}\ }\textbf {\bibinfo {volume} {101}},\ \bibinfo {pages} {045012} (\bibinfo {year} {2020})}\BibitemShut {NoStop}%
\bibitem [{\citenamefont {Fosco}\ \emph {et~al.}(2016)\citenamefont {Fosco}, \citenamefont {Lombardo},\ and\ \citenamefont {Mazzitelli}}]{fosco2016casimir}%
  \BibitemOpen
  \bibfield  {author} {\bibinfo {author} {\bibfnamefont {C.~D.}\ \bibnamefont {Fosco}}, \bibinfo {author} {\bibfnamefont {F.~C.}\ \bibnamefont {Lombardo}},\ and\ \bibinfo {author} {\bibfnamefont {F.~D.}\ \bibnamefont {Mazzitelli}},\ }\href {https://doi.org/10.1103/PhysRevD.93.125015} {\bibfield  {journal} {\bibinfo  {journal} {Physical Review D}\ }\textbf {\bibinfo {volume} {93}},\ \bibinfo {pages} {125015} (\bibinfo {year} {2016})}\BibitemShut {NoStop}%
\bibitem [{\citenamefont {Romaniega}\ \emph {et~al.}(2023)\citenamefont {Romaniega}, \citenamefont {Munoz-Castaneda},\ and\ \citenamefont {Cavero-Pelaez}}]{romaniega2023casimir}%
  \BibitemOpen
  \bibfield  {author} {\bibinfo {author} {\bibfnamefont {C.}~\bibnamefont {Romaniega}}, \bibinfo {author} {\bibfnamefont {J.}~\bibnamefont {Munoz-Castaneda}},\ and\ \bibinfo {author} {\bibfnamefont {I.}~\bibnamefont {Cavero-Pelaez}},\ }\href {https://doi.org/10.1103/PhysRevD.107.025002} {\bibfield  {journal} {\bibinfo  {journal} {Physical Review D}\ }\textbf {\bibinfo {volume} {107}},\ \bibinfo {pages} {025002} (\bibinfo {year} {2023})}\BibitemShut {NoStop}%
\bibitem [{\citenamefont {Cavero-Pel{\'a}ez}\ \emph {et~al.}(2021)\citenamefont {Cavero-Pel{\'a}ez}, \citenamefont {Munoz-Castaneda},\ and\ \citenamefont {Romaniega}}]{cavero2021casimir}%
  \BibitemOpen
  \bibfield  {author} {\bibinfo {author} {\bibfnamefont {I.}~\bibnamefont {Cavero-Pel{\'a}ez}}, \bibinfo {author} {\bibfnamefont {J.~M.}\ \bibnamefont {Munoz-Castaneda}},\ and\ \bibinfo {author} {\bibfnamefont {C.}~\bibnamefont {Romaniega}},\ }\href {https://doi.org/10.1103/PhysRevD.103.045005} {\bibfield  {journal} {\bibinfo  {journal} {Physical Review D}\ }\textbf {\bibinfo {volume} {103}},\ \bibinfo {pages} {045005} (\bibinfo {year} {2021})}\BibitemShut {NoStop}%
\bibitem [{\citenamefont {Shajesh}\ \emph {et~al.}(2016)\citenamefont {Shajesh}, \citenamefont {Brevik}, \citenamefont {Cavero-Pel{\'a}ez},\ and\ \citenamefont {Parashar}}]{shajesh2016casimir}%
  \BibitemOpen
  \bibfield  {author} {\bibinfo {author} {\bibfnamefont {K.~V.}\ \bibnamefont {Shajesh}}, \bibinfo {author} {\bibfnamefont {I.}~\bibnamefont {Brevik}}, \bibinfo {author} {\bibfnamefont {I.}~\bibnamefont {Cavero-Pel{\'a}ez}},\ and\ \bibinfo {author} {\bibfnamefont {P.}~\bibnamefont {Parashar}},\ }\href {https://doi.org/10.1103/PhysRevD.94.065003} {\bibfield  {journal} {\bibinfo  {journal} {Physical Review D}\ }\textbf {\bibinfo {volume} {94}},\ \bibinfo {pages} {065003} (\bibinfo {year} {2016})}\BibitemShut {NoStop}%
\bibitem [{\citenamefont {Shajesh}\ \emph {et~al.}(2017)\citenamefont {Shajesh}, \citenamefont {Parashar}, \citenamefont {Cavero-Pel{\'a}ez}, \citenamefont {Kocik},\ and\ \citenamefont {Brevik}}]{shajesh2017casimir}%
  \BibitemOpen
  \bibfield  {author} {\bibinfo {author} {\bibfnamefont {K.}~\bibnamefont {Shajesh}}, \bibinfo {author} {\bibfnamefont {P.}~\bibnamefont {Parashar}}, \bibinfo {author} {\bibfnamefont {I.}~\bibnamefont {Cavero-Pel{\'a}ez}}, \bibinfo {author} {\bibfnamefont {J.}~\bibnamefont {Kocik}},\ and\ \bibinfo {author} {\bibfnamefont {I.}~\bibnamefont {Brevik}},\ }\href {https://doi.org/10.1103/PhysRevD.96.105010} {\bibfield  {journal} {\bibinfo  {journal} {Physical Review D}\ }\textbf {\bibinfo {volume} {96}},\ \bibinfo {pages} {105010} (\bibinfo {year} {2017})}\BibitemShut {NoStop}%
\bibitem [{\citenamefont {Li}\ \emph {et~al.}(2022)\citenamefont {Li}, \citenamefont {Milton}, \citenamefont {Parashar}, \citenamefont {Kennedy}, \citenamefont {Pourtolami},\ and\ \citenamefont {Guo}}]{li2022casimir}%
  \BibitemOpen
  \bibfield  {author} {\bibinfo {author} {\bibfnamefont {Y.}~\bibnamefont {Li}}, \bibinfo {author} {\bibfnamefont {K.~A.}\ \bibnamefont {Milton}}, \bibinfo {author} {\bibfnamefont {P.}~\bibnamefont {Parashar}}, \bibinfo {author} {\bibfnamefont {G.}~\bibnamefont {Kennedy}}, \bibinfo {author} {\bibfnamefont {N.}~\bibnamefont {Pourtolami}},\ and\ \bibinfo {author} {\bibfnamefont {X.}~\bibnamefont {Guo}},\ }\href {https://doi.org/10.1103/PhysRevD.106.036002} {\bibfield  {journal} {\bibinfo  {journal} {Physical Review D}\ }\textbf {\bibinfo {volume} {106}},\ \bibinfo {pages} {036002} (\bibinfo {year} {2022})}\BibitemShut {NoStop}%
\bibitem [{\citenamefont {Milton}\ \emph {et~al.}(2017)\citenamefont {Milton}, \citenamefont {Kalauni}, \citenamefont {Parashar},\ and\ \citenamefont {Li}}]{milton2017casimir}%
  \BibitemOpen
  \bibfield  {author} {\bibinfo {author} {\bibfnamefont {K.~A.}\ \bibnamefont {Milton}}, \bibinfo {author} {\bibfnamefont {P.}~\bibnamefont {Kalauni}}, \bibinfo {author} {\bibfnamefont {P.}~\bibnamefont {Parashar}},\ and\ \bibinfo {author} {\bibfnamefont {Y.}~\bibnamefont {Li}},\ }\href {https://doi.org/10.1103/PhysRevD.96.085007} {\bibfield  {journal} {\bibinfo  {journal} {Physical Review D}\ }\textbf {\bibinfo {volume} {96}},\ \bibinfo {pages} {085007} (\bibinfo {year} {2017})}\BibitemShut {NoStop}%
\bibitem [{\citenamefont {Milton}\ \emph {et~al.}(2019)\citenamefont {Milton}, \citenamefont {Kalauni}, \citenamefont {Parashar},\ and\ \citenamefont {Li}}]{milton2019remarks}%
  \BibitemOpen
  \bibfield  {author} {\bibinfo {author} {\bibfnamefont {K.~A.}\ \bibnamefont {Milton}}, \bibinfo {author} {\bibfnamefont {P.}~\bibnamefont {Kalauni}}, \bibinfo {author} {\bibfnamefont {P.}~\bibnamefont {Parashar}},\ and\ \bibinfo {author} {\bibfnamefont {Y.}~\bibnamefont {Li}},\ }\href {https://doi.org/10.1103/PhysRevD.99.045013} {\bibfield  {journal} {\bibinfo  {journal} {Physical Review D}\ }\textbf {\bibinfo {volume} {99}},\ \bibinfo {pages} {045013} (\bibinfo {year} {2019})}\BibitemShut {NoStop}%
\bibitem [{\citenamefont {Autumn}\ \emph {et~al.}(2002)\citenamefont {Autumn}, \citenamefont {Sitti}, \citenamefont {Liang}, \citenamefont {Peattie}, \citenamefont {Hansen}, \citenamefont {Sponberg}, \citenamefont {Kenny}, \citenamefont {Fearing}, \citenamefont {Israelachvili},\ and\ \citenamefont {Full}}]{gecko}%
  \BibitemOpen
  \bibfield  {author} {\bibinfo {author} {\bibfnamefont {K.}~\bibnamefont {Autumn}}, \bibinfo {author} {\bibfnamefont {M.}~\bibnamefont {Sitti}}, \bibinfo {author} {\bibfnamefont {Y.~A.}\ \bibnamefont {Liang}}, \bibinfo {author} {\bibfnamefont {A.~M.}\ \bibnamefont {Peattie}}, \bibinfo {author} {\bibfnamefont {W.~R.}\ \bibnamefont {Hansen}}, \bibinfo {author} {\bibfnamefont {S.}~\bibnamefont {Sponberg}}, \bibinfo {author} {\bibfnamefont {T.~W.}\ \bibnamefont {Kenny}}, \bibinfo {author} {\bibfnamefont {R.}~\bibnamefont {Fearing}}, \bibinfo {author} {\bibfnamefont {J.~N.}\ \bibnamefont {Israelachvili}},\ and\ \bibinfo {author} {\bibfnamefont {R.~J.}\ \bibnamefont {Full}},\ }\href@noop {} {\bibfield  {journal} {\bibinfo  {journal} {Proceedings of the National Academy of Sciences}\ }\textbf {\bibinfo {volume} {99}},\ \bibinfo {pages} {12252} (\bibinfo {year} {2002})}\BibitemShut {NoStop}%
\bibitem [{\citenamefont {Lamoreaux}(2007)}]{lamoreaux2007casimir}%
  \BibitemOpen
  \bibfield  {author} {\bibinfo {author} {\bibfnamefont {S.~K.}\ \bibnamefont {Lamoreaux}},\ }\href {https://doi.org/10.1063/1.2711635} {\bibfield  {journal} {\bibinfo  {journal} {Physics Today}\ }\textbf {\bibinfo {volume} {60}},\ \bibinfo {pages} {40} (\bibinfo {year} {2007})}\BibitemShut {NoStop}%
\bibitem [{\citenamefont {Scheeres}\ \emph {et~al.}(2010)\citenamefont {Scheeres}, \citenamefont {Hartzell}, \citenamefont {S{\'a}nchez},\ and\ \citenamefont {Swift}}]{scheeres2010scaling}%
  \BibitemOpen
  \bibfield  {author} {\bibinfo {author} {\bibfnamefont {D.~J.}\ \bibnamefont {Scheeres}}, \bibinfo {author} {\bibfnamefont {C.~M.}\ \bibnamefont {Hartzell}}, \bibinfo {author} {\bibfnamefont {P.}~\bibnamefont {S{\'a}nchez}},\ and\ \bibinfo {author} {\bibfnamefont {M.}~\bibnamefont {Swift}},\ }\href {https://doi.org/j.icarus.2010.07.009} {\bibfield  {journal} {\bibinfo  {journal} {Icarus}\ }\textbf {\bibinfo {volume} {210}},\ \bibinfo {pages} {968} (\bibinfo {year} {2010})}\BibitemShut {NoStop}%
\bibitem [{\citenamefont {Mahanty}\ and\ \citenamefont {Ninham}(1976)}]{mahanty1976}%
  \BibitemOpen
  \bibfield  {author} {\bibinfo {author} {\bibfnamefont {J.}~\bibnamefont {Mahanty}}\ and\ \bibinfo {author} {\bibfnamefont {B.~W.}\ \bibnamefont {Ninham}},\ }\href@noop {} {\emph {\bibinfo {title} {Dispersion Forces}}}\ (\bibinfo  {publisher} {Academic Press},\ \bibinfo {year} {1976})\BibitemShut {NoStop}%
\bibitem [{\citenamefont {Chiao}\ and\ \citenamefont {Boyce}(1999)}]{chiao1999bogoliubov}%
  \BibitemOpen
  \bibfield  {author} {\bibinfo {author} {\bibfnamefont {R.~Y.}\ \bibnamefont {Chiao}}\ and\ \bibinfo {author} {\bibfnamefont {J.}~\bibnamefont {Boyce}},\ }\href {https://doi.org/10.1103/PhysRevA.60.4114} {\bibfield  {journal} {\bibinfo  {journal} {Physical Review A}\ }\textbf {\bibinfo {volume} {60}},\ \bibinfo {pages} {4114} (\bibinfo {year} {1999})}\BibitemShut {NoStop}%
\bibitem [{\citenamefont {de~Melo~e Souza}\ \emph {et~al.}(2015)\citenamefont {de~Melo~e Souza}, \citenamefont {Kort-Kamp}, \citenamefont {Rosa},\ and\ \citenamefont {Farina}}]{Souza2015}%
  \BibitemOpen
  \bibfield  {author} {\bibinfo {author} {\bibfnamefont {R.}~\bibnamefont {de~Melo~e Souza}}, \bibinfo {author} {\bibfnamefont {W.~J.~M.}\ \bibnamefont {Kort-Kamp}}, \bibinfo {author} {\bibfnamefont {F.~S.~S.}\ \bibnamefont {Rosa}},\ and\ \bibinfo {author} {\bibfnamefont {C.}~\bibnamefont {Farina}},\ }\href {https://doi.org/10.1103/PhysRevA.91.052708} {\bibfield  {journal} {\bibinfo  {journal} {Phys.Rev. A}\ }\textbf {\bibinfo {volume} {91}},\ \bibinfo {pages} {052708} (\bibinfo {year} {2015})}\BibitemShut {NoStop}%
\bibitem [{\citenamefont {Grigorio}\ \emph {et~al.}(2012)\citenamefont {Grigorio}, \citenamefont {Guimaraes}, \citenamefont {Rougemont}, \citenamefont {Wotzasek},\ and\ \citenamefont {Zarro}}]{Grigorio:2012jt}%
  \BibitemOpen
  \bibfield  {author} {\bibinfo {author} {\bibfnamefont {L.~S.}\ \bibnamefont {Grigorio}}, \bibinfo {author} {\bibfnamefont {M.~S.}\ \bibnamefont {Guimaraes}}, \bibinfo {author} {\bibfnamefont {R.}~\bibnamefont {Rougemont}}, \bibinfo {author} {\bibfnamefont {C.}~\bibnamefont {Wotzasek}},\ and\ \bibinfo {author} {\bibfnamefont {C.~A.~D.}\ \bibnamefont {Zarro}},\ }\href {https://doi.org/10.1103/PhysRevD.86.027705} {\bibfield  {journal} {\bibinfo  {journal} {Phys. Rev. D}\ }\textbf {\bibinfo {volume} {86}},\ \bibinfo {pages} {027705} (\bibinfo {year} {2012})},\ \Eprint {https://arxiv.org/abs/1202.3798} {arXiv:1202.3798 [hep-th]} \BibitemShut {NoStop}%
\bibitem [{\citenamefont {Guimaraes}\ \emph {et~al.}(2013)\citenamefont {Guimaraes}, \citenamefont {Rougemont}, \citenamefont {Wotzasek},\ and\ \citenamefont {Zarro}}]{Guimaraes:2012tx}%
  \BibitemOpen
  \bibfield  {author} {\bibinfo {author} {\bibfnamefont {M.~S.}\ \bibnamefont {Guimaraes}}, \bibinfo {author} {\bibfnamefont {R.}~\bibnamefont {Rougemont}}, \bibinfo {author} {\bibfnamefont {C.}~\bibnamefont {Wotzasek}},\ and\ \bibinfo {author} {\bibfnamefont {C.~A.~D.}\ \bibnamefont {Zarro}},\ }\href {https://doi.org/10.1016/j.physletb.2013.05.032} {\bibfield  {journal} {\bibinfo  {journal} {Phys. Lett. B}\ }\textbf {\bibinfo {volume} {723}},\ \bibinfo {pages} {422} (\bibinfo {year} {2013})},\ \Eprint {https://arxiv.org/abs/1209.3073} {arXiv:1209.3073 [hep-th]} \BibitemShut {NoStop}%
\bibitem [{\citenamefont {Reinosa}(2024)}]{Reinosa:2024njc}%
  \BibitemOpen
  \bibfield  {author} {\bibinfo {author} {\bibfnamefont {U.}~\bibnamefont {Reinosa}}\ }(\bibinfo {year} {2024})\ \Eprint {https://arxiv.org/abs/2404.06118} {arXiv:2404.06118 [hep-ph]} \BibitemShut {NoStop}%
\bibitem [{\citenamefont {Rougemont}\ \emph {et~al.}(2015)\citenamefont {Rougemont}, \citenamefont {Noronha}, \citenamefont {Zarro}, \citenamefont {Wotzasek}, \citenamefont {Guimaraes},\ and\ \citenamefont {Granado}}]{Rougemont:2015gia}%
  \BibitemOpen
  \bibfield  {author} {\bibinfo {author} {\bibfnamefont {R.}~\bibnamefont {Rougemont}}, \bibinfo {author} {\bibfnamefont {J.}~\bibnamefont {Noronha}}, \bibinfo {author} {\bibfnamefont {C.~A.~D.}\ \bibnamefont {Zarro}}, \bibinfo {author} {\bibfnamefont {C.}~\bibnamefont {Wotzasek}}, \bibinfo {author} {\bibfnamefont {M.~S.}\ \bibnamefont {Guimaraes}},\ and\ \bibinfo {author} {\bibfnamefont {D.~R.}\ \bibnamefont {Granado}},\ }\href {https://doi.org/10.1007/JHEP07(2015)070} {\bibfield  {journal} {\bibinfo  {journal} {JHEP}\ }\textbf {\bibinfo {volume} {07}},\ \bibinfo {pages} {070}},\ \Eprint {https://arxiv.org/abs/1505.02442} {arXiv:1505.02442 [hep-th]} \BibitemShut {NoStop}%
\bibitem [{\citenamefont {Pospelov}(2009)}]{Pospelov:2008zw}%
  \BibitemOpen
  \bibfield  {author} {\bibinfo {author} {\bibfnamefont {M.}~\bibnamefont {Pospelov}},\ }\href {https://doi.org/10.1103/PhysRevD.80.095002} {\bibfield  {journal} {\bibinfo  {journal} {Phys. Rev. D}\ }\textbf {\bibinfo {volume} {80}},\ \bibinfo {pages} {095002} (\bibinfo {year} {2009})},\ \Eprint {https://arxiv.org/abs/0811.1030} {arXiv:0811.1030 [hep-ph]} \BibitemShut {NoStop}%
\bibitem [{\citenamefont {Spallicci}\ \emph {et~al.}(2021)\citenamefont {Spallicci}, \citenamefont {Helay\"el-Neto}, \citenamefont {L\'opez-Corredoira},\ and\ \citenamefont {Capozziello}}]{Spallicci:2020diu}%
  \BibitemOpen
  \bibfield  {author} {\bibinfo {author} {\bibfnamefont {A.~D. A.~M.}\ \bibnamefont {Spallicci}}, \bibinfo {author} {\bibfnamefont {J.~A.}\ \bibnamefont {Helay\"el-Neto}}, \bibinfo {author} {\bibfnamefont {M.}~\bibnamefont {L\'opez-Corredoira}},\ and\ \bibinfo {author} {\bibfnamefont {S.}~\bibnamefont {Capozziello}},\ }\href {https://doi.org/10.1140/epjc/s10052-020-08703-3} {\bibfield  {journal} {\bibinfo  {journal} {Eur. Phys. J. C}\ }\textbf {\bibinfo {volume} {81}},\ \bibinfo {pages} {4} (\bibinfo {year} {2021})},\ \Eprint {https://arxiv.org/abs/2011.12608} {arXiv:2011.12608 [astro-ph.CO]} \BibitemShut {NoStop}%
\bibitem [{\citenamefont {Bordag}\ \emph {et~al.}(1998)\citenamefont {Bordag}, \citenamefont {Geyer}, \citenamefont {Klimchitskaya},\ and\ \citenamefont {Mostepanenko}}]{bordag1998constraints}%
  \BibitemOpen
  \bibfield  {author} {\bibinfo {author} {\bibfnamefont {M.}~\bibnamefont {Bordag}}, \bibinfo {author} {\bibfnamefont {B.}~\bibnamefont {Geyer}}, \bibinfo {author} {\bibfnamefont {G.}~\bibnamefont {Klimchitskaya}},\ and\ \bibinfo {author} {\bibfnamefont {V.}~\bibnamefont {Mostepanenko}},\ }\href {https://doi.org/10.1103/PhysRevD.58.075003} {\bibfield  {journal} {\bibinfo  {journal} {Physical Review D}\ }\textbf {\bibinfo {volume} {58}},\ \bibinfo {pages} {075003} (\bibinfo {year} {1998})}\BibitemShut {NoStop}%
\bibitem [{\citenamefont {Fischbach}\ \emph {et~al.}(2001)\citenamefont {Fischbach}, \citenamefont {Krause}, \citenamefont {Mostepanenko},\ and\ \citenamefont {Novello}}]{fischbach2001new}%
  \BibitemOpen
  \bibfield  {author} {\bibinfo {author} {\bibfnamefont {E.}~\bibnamefont {Fischbach}}, \bibinfo {author} {\bibfnamefont {D.}~\bibnamefont {Krause}}, \bibinfo {author} {\bibfnamefont {V.}~\bibnamefont {Mostepanenko}},\ and\ \bibinfo {author} {\bibfnamefont {M.}~\bibnamefont {Novello}},\ }\href {https://doi.org/10.1103/PhysRevD.64.075010} {\bibfield  {journal} {\bibinfo  {journal} {Physical Review D}\ }\textbf {\bibinfo {volume} {64}},\ \bibinfo {pages} {075010} (\bibinfo {year} {2001})}\BibitemShut {NoStop}%
\bibitem [{\citenamefont {Mostepanenko}(2016)}]{mostepanenko2016progress}%
  \BibitemOpen
  \bibfield  {author} {\bibinfo {author} {\bibfnamefont {V.}~\bibnamefont {Mostepanenko}},\ }\href {https://doi.org/10.1142/S0217751X16410207} {\bibfield  {journal} {\bibinfo  {journal} {International Journal of Modern Physics A}\ }\textbf {\bibinfo {volume} {31}},\ \bibinfo {pages} {1641020} (\bibinfo {year} {2016})}\BibitemShut {NoStop}%
\bibitem [{\citenamefont {Adelberger}\ \emph {et~al.}(2007)\citenamefont {Adelberger}, \citenamefont {Dvali},\ and\ \citenamefont {Gruzinov}}]{adelberger2007photon}%
  \BibitemOpen
  \bibfield  {author} {\bibinfo {author} {\bibfnamefont {E.}~\bibnamefont {Adelberger}}, \bibinfo {author} {\bibfnamefont {G.}~\bibnamefont {Dvali}},\ and\ \bibinfo {author} {\bibfnamefont {A.}~\bibnamefont {Gruzinov}},\ }\href {https://doi.org/10.1103/PhysRevLett.98.010402} {\bibfield  {journal} {\bibinfo  {journal} {Physical Review Letters}\ }\textbf {\bibinfo {volume} {98}},\ \bibinfo {pages} {010402} (\bibinfo {year} {2007})}\BibitemShut {NoStop}%
\bibitem [{\citenamefont {Chibisov}(1976)}]{chibisov1976astrophysical}%
  \BibitemOpen
  \bibfield  {author} {\bibinfo {author} {\bibfnamefont {G.~V.}\ \bibnamefont {Chibisov}},\ }\href {https://doi.org/10.1070/PU1976v019n07ABEH005277} {\bibfield  {journal} {\bibinfo  {journal} {Soviet Physics Uspekhi}\ }\textbf {\bibinfo {volume} {19}},\ \bibinfo {pages} {624} (\bibinfo {year} {1976})}\BibitemShut {NoStop}%
\bibitem [{\citenamefont {Accioly}\ \emph {et~al.}(2010)\citenamefont {Accioly}, \citenamefont {Helay{\"e}l-Neto},\ and\ \citenamefont {Scatena}}]{accioly2010upper}%
  \BibitemOpen
  \bibfield  {author} {\bibinfo {author} {\bibfnamefont {A.}~\bibnamefont {Accioly}}, \bibinfo {author} {\bibfnamefont {J.}~\bibnamefont {Helay{\"e}l-Neto}},\ and\ \bibinfo {author} {\bibfnamefont {E.}~\bibnamefont {Scatena}},\ }\href {https://doi.org/10.1103/PhysRevD.82.065026} {\bibfield  {journal} {\bibinfo  {journal} {Physical Review D}\ }\textbf {\bibinfo {volume} {82}},\ \bibinfo {pages} {065026} (\bibinfo {year} {2010})}\BibitemShut {NoStop}%
\bibitem [{\citenamefont {Malta}\ and\ \citenamefont {Helay{\"e}l-Neto}(2022)}]{malta2022constraining}%
  \BibitemOpen
  \bibfield  {author} {\bibinfo {author} {\bibfnamefont {P.}~\bibnamefont {Malta}}\ and\ \bibinfo {author} {\bibfnamefont {J.}~\bibnamefont {Helay{\"e}l-Neto}},\ }\href {https://doi.org/10.1103/PhysRevD.106.116014} {\bibfield  {journal} {\bibinfo  {journal} {Physical Review D}\ }\textbf {\bibinfo {volume} {106}},\ \bibinfo {pages} {116014} (\bibinfo {year} {2022})}\BibitemShut {NoStop}%
\bibitem [{\citenamefont {Malta}(2023)}]{malta2023shining}%
  \BibitemOpen
  \bibfield  {author} {\bibinfo {author} {\bibfnamefont {P.}~\bibnamefont {Malta}},\ }\href {https://doi.org/10.1103/PhysRevD.107.115018} {\bibfield  {journal} {\bibinfo  {journal} {Physical Review D}\ }\textbf {\bibinfo {volume} {107}},\ \bibinfo {pages} {115018} (\bibinfo {year} {2023})}\BibitemShut {NoStop}%
\bibitem [{\citenamefont {Goldhaber}\ and\ \citenamefont {Nieto}(1971)}]{goldhaber1971terrestrial}%
  \BibitemOpen
  \bibfield  {author} {\bibinfo {author} {\bibfnamefont {A.~S.}\ \bibnamefont {Goldhaber}}\ and\ \bibinfo {author} {\bibfnamefont {M.~M.}\ \bibnamefont {Nieto}},\ }\href {https://doi.org/10.1103/RevModPhys.43.277} {\bibfield  {journal} {\bibinfo  {journal} {Reviews of Modern Physics}\ }\textbf {\bibinfo {volume} {43}},\ \bibinfo {pages} {277} (\bibinfo {year} {1971})}\BibitemShut {NoStop}%
\bibitem [{\citenamefont {Goldhaber}\ and\ \citenamefont {Nieto}(2010)}]{goldhaber2010photon}%
  \BibitemOpen
  \bibfield  {author} {\bibinfo {author} {\bibfnamefont {A.~S.}\ \bibnamefont {Goldhaber}}\ and\ \bibinfo {author} {\bibfnamefont {M.~M.}\ \bibnamefont {Nieto}},\ }\href {https://doi.org/10.1103/RevModPhys.82.939} {\bibfield  {journal} {\bibinfo  {journal} {Reviews of Modern Physics}\ }\textbf {\bibinfo {volume} {82}},\ \bibinfo {pages} {939} (\bibinfo {year} {2010})}\BibitemShut {NoStop}%
\bibitem [{\citenamefont {Glendenning}(1997)}]{norman1997compact}%
  \BibitemOpen
  \bibfield  {author} {\bibinfo {author} {\bibfnamefont {N.~K.}\ \bibnamefont {Glendenning}},\ }\href {https://books.google.com.br/books?id=57XvAAAAMAAJ} {\emph {\bibinfo {title} {Compact Stars: Nuclear Physics, Particle Physics and General Relativity}}},\ Astronomy and Astrophysics Library\ (\bibinfo  {publisher} {Springer New York},\ \bibinfo {year} {1997})\BibitemShut {NoStop}%
\bibitem [{\citenamefont {Horowitz}\ and\ \citenamefont {Piekarewicz}(2001)}]{Horowitz:2000xj}%
  \BibitemOpen
  \bibfield  {author} {\bibinfo {author} {\bibfnamefont {C.~J.}\ \bibnamefont {Horowitz}}\ and\ \bibinfo {author} {\bibfnamefont {J.}~\bibnamefont {Piekarewicz}},\ }\href {https://doi.org/10.1103/PhysRevLett.86.5647} {\bibfield  {journal} {\bibinfo  {journal} {Phys. Rev. Lett.}\ }\textbf {\bibinfo {volume} {86}},\ \bibinfo {pages} {5647} (\bibinfo {year} {2001})},\ \Eprint {https://arxiv.org/abs/astro-ph/0010227} {arXiv:astro-ph/0010227} \BibitemShut {NoStop}%
\bibitem [{\citenamefont {Chen}\ and\ \citenamefont {Piekarewicz}(2014)}]{Chen:2014sca}%
  \BibitemOpen
  \bibfield  {author} {\bibinfo {author} {\bibfnamefont {W.-C.}\ \bibnamefont {Chen}}\ and\ \bibinfo {author} {\bibfnamefont {J.}~\bibnamefont {Piekarewicz}},\ }\href {https://doi.org/10.1103/PhysRevC.90.044305} {\bibfield  {journal} {\bibinfo  {journal} {Phys. Rev. C}\ }\textbf {\bibinfo {volume} {90}},\ \bibinfo {pages} {044305} (\bibinfo {year} {2014})},\ \Eprint {https://arxiv.org/abs/1408.4159} {arXiv:1408.4159 [nucl-th]} \BibitemShut {NoStop}%
\bibitem [{\citenamefont {Walecka}(1974)}]{Walecka:1974qa}%
  \BibitemOpen
  \bibfield  {author} {\bibinfo {author} {\bibfnamefont {J.~D.}\ \bibnamefont {Walecka}},\ }\href {https://doi.org/10.1016/0003-4916(74)90208-5} {\bibfield  {journal} {\bibinfo  {journal} {Annals Phys.}\ }\textbf {\bibinfo {volume} {83}},\ \bibinfo {pages} {491} (\bibinfo {year} {1974})}\BibitemShut {NoStop}%
\bibitem [{\citenamefont {Boguta}(1983)}]{Boguta:1982wr}%
  \BibitemOpen
  \bibfield  {author} {\bibinfo {author} {\bibfnamefont {J.}~\bibnamefont {Boguta}},\ }\href {https://doi.org/10.1016/0370-2693(83)90617-2} {\bibfield  {journal} {\bibinfo  {journal} {Phys. Lett. B}\ }\textbf {\bibinfo {volume} {120}},\ \bibinfo {pages} {34} (\bibinfo {year} {1983})}\BibitemShut {NoStop}%
\bibitem [{\citenamefont {Pisarski}(2021)}]{Pisarski:2021aoz}%
  \BibitemOpen
  \bibfield  {author} {\bibinfo {author} {\bibfnamefont {R.~D.}\ \bibnamefont {Pisarski}},\ }\href {https://doi.org/10.1103/PhysRevD.103.L071504} {\bibfield  {journal} {\bibinfo  {journal} {Phys. Rev. D}\ }\textbf {\bibinfo {volume} {103}},\ \bibinfo {pages} {L071504} (\bibinfo {year} {2021})},\ \Eprint {https://arxiv.org/abs/2101.05813} {arXiv:2101.05813 [nucl-th]} \BibitemShut {NoStop}%
\bibitem [{\citenamefont {Willey}(1978)}]{willey1978quark}%
  \BibitemOpen
  \bibfield  {author} {\bibinfo {author} {\bibfnamefont {R.}~\bibnamefont {Willey}},\ }\href {https://doi.org/10.1103/PhysRevD.18.270} {\bibfield  {journal} {\bibinfo  {journal} {Physical Review D}\ }\textbf {\bibinfo {volume} {18}},\ \bibinfo {pages} {270} (\bibinfo {year} {1978})}\BibitemShut {NoStop}%
\bibitem [{\citenamefont {Fujii}\ and\ \citenamefont {Kharzeev}(1999)}]{Fujii:1999xn}%
  \BibitemOpen
  \bibfield  {author} {\bibinfo {author} {\bibfnamefont {H.}~\bibnamefont {Fujii}}\ and\ \bibinfo {author} {\bibfnamefont {D.}~\bibnamefont {Kharzeev}},\ }\href {https://doi.org/10.1103/PhysRevD.60.114039} {\bibfield  {journal} {\bibinfo  {journal} {Phys. Rev. D}\ }\textbf {\bibinfo {volume} {60}},\ \bibinfo {pages} {114039} (\bibinfo {year} {1999})},\ \Eprint {https://arxiv.org/abs/hep-ph/9903495} {arXiv:hep-ph/9903495} \BibitemShut {NoStop}%
\bibitem [{\citenamefont {Proca}(1936)}]{proca1936theorie}%
  \BibitemOpen
  \bibfield  {author} {\bibinfo {author} {\bibfnamefont {A.}~\bibnamefont {Proca}},\ }\href {https://doi.org/10.1051/jphysrad:0193600708034700} {\bibfield  {journal} {\bibinfo  {journal} {Journal de Physique et le Radium}\ }\textbf {\bibinfo {volume} {7}},\ \bibinfo {pages} {347} (\bibinfo {year} {1936})}\BibitemShut {NoStop}%
\bibitem [{\citenamefont {Davies}\ and\ \citenamefont {Unwin}(1981)}]{daviesPLB}%
  \BibitemOpen
  \bibfield  {author} {\bibinfo {author} {\bibfnamefont {P.}~\bibnamefont {Davies}}\ and\ \bibinfo {author} {\bibfnamefont {S.}~\bibnamefont {Unwin}},\ }\href {https://doi.org/10.1016/0370-2693(81)90013-7} {\bibfield  {journal} {\bibinfo  {journal} {Physical Letters B}\ }\textbf {\bibinfo {volume} {98}},\ \bibinfo {pages} {274} (\bibinfo {year} {1981})}\BibitemShut {NoStop}%
\bibitem [{\citenamefont {Barton}\ and\ \citenamefont {Dombey}(1984)}]{barton1984casimir}%
  \BibitemOpen
  \bibfield  {author} {\bibinfo {author} {\bibfnamefont {G.}~\bibnamefont {Barton}}\ and\ \bibinfo {author} {\bibfnamefont {N.}~\bibnamefont {Dombey}},\ }\href {https://doi.org/10.1038/311336a0} {\bibfield  {journal} {\bibinfo  {journal} {Nature}\ }\textbf {\bibinfo {volume} {311}},\ \bibinfo {pages} {336} (\bibinfo {year} {1984})}\BibitemShut {NoStop}%
\bibitem [{\citenamefont {Barton}\ and\ \citenamefont {Dombey}(1985)}]{barton1985casimir}%
  \BibitemOpen
  \bibfield  {author} {\bibinfo {author} {\bibfnamefont {G.}~\bibnamefont {Barton}}\ and\ \bibinfo {author} {\bibfnamefont {N.}~\bibnamefont {Dombey}},\ }\href {https://doi.org/https://doi.org/10.1016/0003-4916(85)90162-9} {\bibfield  {journal} {\bibinfo  {journal} {Annals of Physics}\ }\textbf {\bibinfo {volume} {162}},\ \bibinfo {pages} {231} (\bibinfo {year} {1985})}\BibitemShut {NoStop}%
\bibitem [{\citenamefont {Mattioli}\ \emph {et~al.}(2019)\citenamefont {Mattioli}, \citenamefont {Frassino},\ and\ \citenamefont {Panella}}]{mattioli2019casimir}%
  \BibitemOpen
  \bibfield  {author} {\bibinfo {author} {\bibfnamefont {L.}~\bibnamefont {Mattioli}}, \bibinfo {author} {\bibfnamefont {A.}~\bibnamefont {Frassino}},\ and\ \bibinfo {author} {\bibfnamefont {O.}~\bibnamefont {Panella}},\ }\href {https://doi.org/10.1103/PhysRevD.100.116023} {\bibfield  {journal} {\bibinfo  {journal} {Physical Review D}\ }\textbf {\bibinfo {volume} {100}},\ \bibinfo {pages} {116023} (\bibinfo {year} {2019})}\BibitemShut {NoStop}%
\bibitem [{\citenamefont {Craig}\ and\ \citenamefont {Thirunamachandran}(1998)}]{craig1998molecular}%
  \BibitemOpen
  \bibfield  {author} {\bibinfo {author} {\bibfnamefont {D.~P.}\ \bibnamefont {Craig}}\ and\ \bibinfo {author} {\bibfnamefont {T.}~\bibnamefont {Thirunamachandran}},\ }\href@noop {} {\emph {\bibinfo {title} {Molecular quantum electrodynamics: an introduction to radiation-molecule interactions}}}\ (\bibinfo  {publisher} {Courier Corporation},\ \bibinfo {year} {1998})\BibitemShut {NoStop}%
\bibitem [{\citenamefont {Greiner}\ and\ \citenamefont {Reinhardt}(1996)}]{greiner1996field}%
  \BibitemOpen
  \bibfield  {author} {\bibinfo {author} {\bibfnamefont {W.}~\bibnamefont {Greiner}}\ and\ \bibinfo {author} {\bibfnamefont {J.}~\bibnamefont {Reinhardt}},\ }\href@noop {} {\emph {\bibinfo {title} {Field quantization}}}\ (\bibinfo  {publisher} {Springer Science \& Business Media},\ \bibinfo {year} {1996})\BibitemShut {NoStop}%
\bibitem [{\citenamefont {Casimir}\ and\ \citenamefont {Polder}(1948)}]{casimir1948influence}%
  \BibitemOpen
  \bibfield  {author} {\bibinfo {author} {\bibfnamefont {H.~B.~G.}\ \bibnamefont {Casimir}}\ and\ \bibinfo {author} {\bibfnamefont {D.}~\bibnamefont {Polder}},\ }\href {https://doi.org/10.1103/PhysRev.73.360} {\bibfield  {journal} {\bibinfo  {journal} {Physical Review}\ }\textbf {\bibinfo {volume} {73}},\ \bibinfo {pages} {360} (\bibinfo {year} {1948})}\BibitemShut {NoStop}%
\bibitem [{\citenamefont {Peskin}\ and\ \citenamefont {Schroeder}(1995)}]{peskin}%
  \BibitemOpen
  \bibfield  {author} {\bibinfo {author} {\bibfnamefont {M.~E.}\ \bibnamefont {Peskin}}\ and\ \bibinfo {author} {\bibfnamefont {D.~V.}\ \bibnamefont {Schroeder}},\ }\href@noop {} {\emph {\bibinfo {title} {{An Introduction to quantum field theory}}}}\ (\bibinfo  {publisher} {Addison-Wesley},\ \bibinfo {address} {Reading, USA},\ \bibinfo {year} {1995})\BibitemShut {NoStop}%
\bibitem [{Notei()}]{Notei}%
  \BibitemOpen
  \bibinfo {note} {We note that the asymptotic expressions for $\mu R\gg 1$ have different behaviors for the two approximations. However, this difference only becomes pronounced for small values of $U(\mu R)/U(\mu R =0)$, which explains why they are not visible in Fig.~\ref {band}.}\BibitemShut {Stop}%
\bibitem [{\citenamefont {Dragulin}\ \emph {et~al.}(2008)\citenamefont {Dragulin}, \citenamefont {Leung} \emph {et~al.}}]{dragulin2008green}%
  \BibitemOpen
  \bibfield  {author} {\bibinfo {author} {\bibfnamefont {P.}~\bibnamefont {Dragulin}}, \bibinfo {author} {\bibfnamefont {P.}~\bibnamefont {Leung}}, \emph {et~al.},\ }\href {https://doi.org/10.1103/PhysRevE.78.026605} {\bibfield  {journal} {\bibinfo  {journal} {Physical Review E}\ }\textbf {\bibinfo {volume} {78}},\ \bibinfo {pages} {026605} (\bibinfo {year} {2008})}\BibitemShut {NoStop}%
\bibitem [{\citenamefont {Pitombo}\ \emph {et~al.}(2021)\citenamefont {Pitombo}, \citenamefont {Vasconcellos}, \citenamefont {Farina},\ and\ \citenamefont {de~Melo~e Souza}}]{Pitombo2021}%
  \BibitemOpen
  \bibfield  {author} {\bibinfo {author} {\bibfnamefont {R.~S.}\ \bibnamefont {Pitombo}}, \bibinfo {author} {\bibfnamefont {M.}~\bibnamefont {Vasconcellos}}, \bibinfo {author} {\bibfnamefont {C.}~\bibnamefont {Farina}},\ and\ \bibinfo {author} {\bibfnamefont {R.}~\bibnamefont {de~Melo~e Souza}},\ }\href {https://doi.org/10.1088/1361-6404/abcba5} {\bibfield  {journal} {\bibinfo  {journal} {European Journal of Physics}\ }\textbf {\bibinfo {volume} {42}},\ \bibinfo {pages} {025202} (\bibinfo {year} {2021})}\BibitemShut {NoStop}%
\end{thebibliography}

%

\end{document}